\documentclass[letterpaper,twocolumn,10pt]{article}
\usepackage{sty/usenix2019_v3.1}

\usepackage{tikz}
\usepackage{amsmath}
\usepackage{filecontents}

\usepackage{amsmath,amsopn}
\usepackage{subfigure}
\usepackage{endnotes}
\usepackage{microtype}
\usepackage{xspace}
\usepackage{graphicx}
\usepackage{fancyvrb}
\usepackage{multirow}
\usepackage{array}
\usepackage{underscore}
\usepackage{relsize}
\usepackage{pbox}
\usepackage{xstring}
\usepackage{tikz}
\usepackage[ruled,vlined]{algorithm2e}
\usepackage{fp}
\usepackage{siunitx}
\usepackage{flushend}
\usepackage{balance}
\usepackage{multicol}
\usepackage{hyperref}
\usepackage{booktabs}
\usepackage{makecell}
\usepackage{enumitem}
\usepackage{color, colortbl}
\usepackage{float}
\usepackage{placeins}
\usepackage{stfloats}
\usepackage{authblk}

\makeatletter
\def\mdseries@tt{m}
\makeatother
\usepackage[frozencache=true,cachedir=minted-cache]{minted}
\usepackage{filecontents}

\fvset{fontsize=\scriptsize,xleftmargin=8pt,numbers=left,numbersep=5pt}

\makeatletter
\def\PY@reset{\let\PY@it=\relax \let\PY@bf=\relax%
    \let\PY@ul=\relax \let\PY@tc=\relax%
    \let\PY@bc=\relax \let\PY@ff=\relax}
\def\PY@tok#1{\csname PY@tok@#1\endcsname}
\def\PY@toks#1+{\ifx\relax#1\empty\else%
    \PY@tok{#1}\expandafter\PY@toks\fi}
\def\PY@do#1{\PY@bc{\PY@tc{\PY@ul{%
    \PY@it{\PY@bf{\PY@ff{#1}}}}}}}
\def\PY#1#2{\PY@reset\PY@toks#1+\relax+\PY@do{#2}}

\expandafter\def\csname PY@tok@gd\endcsname{\def\PY@tc##1{\textcolor[rgb]{0.63,0.00,0.00}{##1}}}
\expandafter\def\csname PY@tok@gu\endcsname{\let\PY@bf=\textbf\def\PY@tc##1{\textcolor[rgb]{0.50,0.00,0.50}{##1}}}
\expandafter\def\csname PY@tok@gt\endcsname{\def\PY@tc##1{\textcolor[rgb]{0.00,0.27,0.87}{##1}}}
\expandafter\def\csname PY@tok@gs\endcsname{\let\PY@bf=\textbf}
\expandafter\def\csname PY@tok@gr\endcsname{\def\PY@tc##1{\textcolor[rgb]{1.00,0.00,0.00}{##1}}}
\expandafter\def\csname PY@tok@cm\endcsname{\let\PY@it=\textit\def\PY@tc##1{\textcolor[rgb]{0.25,0.50,0.50}{##1}}}
\expandafter\def\csname PY@tok@vg\endcsname{\def\PY@tc##1{\textcolor[rgb]{0.10,0.09,0.49}{##1}}}
\expandafter\def\csname PY@tok@vi\endcsname{\def\PY@tc##1{\textcolor[rgb]{0.10,0.09,0.49}{##1}}}
\expandafter\def\csname PY@tok@vm\endcsname{\def\PY@tc##1{\textcolor[rgb]{0.10,0.09,0.49}{##1}}}
\expandafter\def\csname PY@tok@mh\endcsname{\def\PY@tc##1{\textcolor[rgb]{0.40,0.40,0.40}{##1}}}
\expandafter\def\csname PY@tok@cs\endcsname{\let\PY@it=\textit\def\PY@tc##1{\textcolor[rgb]{0.25,0.50,0.50}{##1}}}
\expandafter\def\csname PY@tok@ge\endcsname{\let\PY@it=\textit}
\expandafter\def\csname PY@tok@vc\endcsname{\def\PY@tc##1{\textcolor[rgb]{0.10,0.09,0.49}{##1}}}
\expandafter\def\csname PY@tok@il\endcsname{\def\PY@tc##1{\textcolor[rgb]{0.40,0.40,0.40}{##1}}}
\expandafter\def\csname PY@tok@go\endcsname{\def\PY@tc##1{\textcolor[rgb]{0.53,0.53,0.53}{##1}}}
\expandafter\def\csname PY@tok@cp\endcsname{\def\PY@tc##1{\textcolor[rgb]{0.74,0.48,0.00}{##1}}}
\expandafter\def\csname PY@tok@gi\endcsname{\def\PY@tc##1{\textcolor[rgb]{0.00,0.63,0.00}{##1}}}
\expandafter\def\csname PY@tok@gh\endcsname{\let\PY@bf=\textbf\def\PY@tc##1{\textcolor[rgb]{0.00,0.00,0.50}{##1}}}
\expandafter\def\csname PY@tok@ni\endcsname{\let\PY@bf=\textbf\def\PY@tc##1{\textcolor[rgb]{0.60,0.60,0.60}{##1}}}
\expandafter\def\csname PY@tok@nl\endcsname{\def\PY@tc##1{\textcolor[rgb]{0.63,0.63,0.00}{##1}}}
\expandafter\def\csname PY@tok@nn\endcsname{\let\PY@bf=\textbf\def\PY@tc##1{\textcolor[rgb]{0.00,0.00,1.00}{##1}}}
\expandafter\def\csname PY@tok@no\endcsname{\def\PY@tc##1{\textcolor[rgb]{0.53,0.00,0.00}{##1}}}
\expandafter\def\csname PY@tok@na\endcsname{\def\PY@tc##1{\textcolor[rgb]{0.49,0.56,0.16}{##1}}}
\expandafter\def\csname PY@tok@nb\endcsname{\def\PY@tc##1{\textcolor[rgb]{0.00,0.50,0.00}{##1}}}
\expandafter\def\csname PY@tok@nc\endcsname{\let\PY@bf=\textbf\def\PY@tc##1{\textcolor[rgb]{0.00,0.00,1.00}{##1}}}
\expandafter\def\csname PY@tok@nd\endcsname{\def\PY@tc##1{\textcolor[rgb]{0.67,0.13,1.00}{##1}}}
\expandafter\def\csname PY@tok@ne\endcsname{\let\PY@bf=\textbf\def\PY@tc##1{\textcolor[rgb]{0.82,0.25,0.23}{##1}}}
\expandafter\def\csname PY@tok@nf\endcsname{\def\PY@tc##1{\textcolor[rgb]{0.00,0.00,1.00}{##1}}}
\expandafter\def\csname PY@tok@si\endcsname{\let\PY@bf=\textbf\def\PY@tc##1{\textcolor[rgb]{0.73,0.40,0.53}{##1}}}
\expandafter\def\csname PY@tok@s2\endcsname{\def\PY@tc##1{\textcolor[rgb]{0.73,0.13,0.13}{##1}}}
\expandafter\def\csname PY@tok@nt\endcsname{\let\PY@bf=\textbf\def\PY@tc##1{\textcolor[rgb]{0.00,0.50,0.00}{##1}}}
\expandafter\def\csname PY@tok@nv\endcsname{\def\PY@tc##1{\textcolor[rgb]{0.10,0.09,0.49}{##1}}}
\expandafter\def\csname PY@tok@s1\endcsname{\def\PY@tc##1{\textcolor[rgb]{0.73,0.13,0.13}{##1}}}
\expandafter\def\csname PY@tok@dl\endcsname{\def\PY@tc##1{\textcolor[rgb]{0.73,0.13,0.13}{##1}}}
\expandafter\def\csname PY@tok@ch\endcsname{\let\PY@it=\textit\def\PY@tc##1{\textcolor[rgb]{0.25,0.50,0.50}{##1}}}
\expandafter\def\csname PY@tok@m\endcsname{\def\PY@tc##1{\textcolor[rgb]{0.40,0.40,0.40}{##1}}}
\expandafter\def\csname PY@tok@gp\endcsname{\let\PY@bf=\textbf\def\PY@tc##1{\textcolor[rgb]{0.00,0.00,0.50}{##1}}}
\expandafter\def\csname PY@tok@sh\endcsname{\def\PY@tc##1{\textcolor[rgb]{0.73,0.13,0.13}{##1}}}
\expandafter\def\csname PY@tok@ow\endcsname{\let\PY@bf=\textbf\def\PY@tc##1{\textcolor[rgb]{0.67,0.13,1.00}{##1}}}
\expandafter\def\csname PY@tok@sx\endcsname{\def\PY@tc##1{\textcolor[rgb]{0.00,0.50,0.00}{##1}}}
\expandafter\def\csname PY@tok@bp\endcsname{\def\PY@tc##1{\textcolor[rgb]{0.00,0.50,0.00}{##1}}}
\expandafter\def\csname PY@tok@c1\endcsname{\let\PY@it=\textit\def\PY@tc##1{\textcolor[rgb]{0.25,0.50,0.50}{##1}}}
\expandafter\def\csname PY@tok@fm\endcsname{\def\PY@tc##1{\textcolor[rgb]{0.00,0.00,1.00}{##1}}}
\expandafter\def\csname PY@tok@o\endcsname{\def\PY@tc##1{\textcolor[rgb]{0.40,0.40,0.40}{##1}}}
\expandafter\def\csname PY@tok@kc\endcsname{\let\PY@bf=\textbf\def\PY@tc##1{\textcolor[rgb]{0.00,0.50,0.00}{##1}}}
\expandafter\def\csname PY@tok@c\endcsname{\let\PY@it=\textit\def\PY@tc##1{\textcolor[rgb]{0.25,0.50,0.50}{##1}}}
\expandafter\def\csname PY@tok@mf\endcsname{\def\PY@tc##1{\textcolor[rgb]{0.40,0.40,0.40}{##1}}}
\expandafter\def\csname PY@tok@err\endcsname{\def\PY@bc##1{\setlength{\fboxsep}{0pt}\fcolorbox[rgb]{1.00,0.00,0.00}{1,1,1}{\strut ##1}}}
\expandafter\def\csname PY@tok@mb\endcsname{\def\PY@tc##1{\textcolor[rgb]{0.40,0.40,0.40}{##1}}}
\expandafter\def\csname PY@tok@ss\endcsname{\def\PY@tc##1{\textcolor[rgb]{0.10,0.09,0.49}{##1}}}
\expandafter\def\csname PY@tok@sr\endcsname{\def\PY@tc##1{\textcolor[rgb]{0.73,0.40,0.53}{##1}}}
\expandafter\def\csname PY@tok@mo\endcsname{\def\PY@tc##1{\textcolor[rgb]{0.40,0.40,0.40}{##1}}}
\expandafter\def\csname PY@tok@kd\endcsname{\let\PY@bf=\textbf\def\PY@tc##1{\textcolor[rgb]{0.00,0.50,0.00}{##1}}}
\expandafter\def\csname PY@tok@mi\endcsname{\def\PY@tc##1{\textcolor[rgb]{0.40,0.40,0.40}{##1}}}
\expandafter\def\csname PY@tok@kn\endcsname{\let\PY@bf=\textbf\def\PY@tc##1{\textcolor[rgb]{0.00,0.50,0.00}{##1}}}
\expandafter\def\csname PY@tok@cpf\endcsname{\let\PY@it=\textit\def\PY@tc##1{\textcolor[rgb]{0.25,0.50,0.50}{##1}}}
\expandafter\def\csname PY@tok@kr\endcsname{\let\PY@bf=\textbf\def\PY@tc##1{\textcolor[rgb]{0.00,0.50,0.00}{##1}}}
\expandafter\def\csname PY@tok@s\endcsname{\def\PY@tc##1{\textcolor[rgb]{0.73,0.13,0.13}{##1}}}
\expandafter\def\csname PY@tok@kp\endcsname{\def\PY@tc##1{\textcolor[rgb]{0.00,0.50,0.00}{##1}}}
\expandafter\def\csname PY@tok@w\endcsname{\def\PY@tc##1{\textcolor[rgb]{0.73,0.73,0.73}{##1}}}
\expandafter\def\csname PY@tok@kt\endcsname{\def\PY@tc##1{\textcolor[rgb]{0.69,0.00,0.25}{##1}}}
\expandafter\def\csname PY@tok@sc\endcsname{\def\PY@tc##1{\textcolor[rgb]{0.73,0.13,0.13}{##1}}}
\expandafter\def\csname PY@tok@sb\endcsname{\def\PY@tc##1{\textcolor[rgb]{0.73,0.13,0.13}{##1}}}
\expandafter\def\csname PY@tok@sa\endcsname{\def\PY@tc##1{\textcolor[rgb]{0.73,0.13,0.13}{##1}}}
\expandafter\def\csname PY@tok@k\endcsname{\let\PY@bf=\textbf\def\PY@tc##1{\textcolor[rgb]{0.00,0.50,0.00}{##1}}}
\expandafter\def\csname PY@tok@se\endcsname{\let\PY@bf=\textbf\def\PY@tc##1{\textcolor[rgb]{0.73,0.40,0.13}{##1}}}
\expandafter\def\csname PY@tok@sd\endcsname{\let\PY@it=\textit\def\PY@tc##1{\textcolor[rgb]{0.73,0.13,0.13}{##1}}}

\makeatother

\def\Snospace~{\S{}}

\newcommand{\PP}[1]{
\vspace{2px}
{\bf \IfEndWith{#1}{.}{#1}{#1.}}
}

\usepackage{pifont}
\newcommand{\cmark}{\ding{51}}%
\newcommand{\xmark}{\ding{55}}%

\definecolor{Gray}{gray}{0.9}

\newcommand{\boxbeg}{
\vspace{2px}
\noindent\begin{tabular}{|l|}\hline
\begin{minipage}{3.2in}
\vspace{2px}
\noindent
}

\newcommand{\boxend}{
\vspace{2px}
\end{minipage}\\ \hline
\end{tabular}
\vspace{-10pt}
}

\newcommand{\gitrepo}{\textit{\url{https://github.com/Jamrot/ChatGPT-Vulnerability-Management}}\xspace}

\newcommand{\gptthree}{gpt-3.5\xspace}
\newcommand{\gptfour}{gpt-4\xspace}

\newcommand{\zeroshot}{0-shot\xspace}
\newcommand{\oneshot}{1-shot\xspace}
\newcommand{\fewshot}{few-shot\xspace}
\newcommand{\prompteng}{general-info\xspace}
\newcommand{\manualinfo}{expertise\xspace}
\newcommand{\gptinfo}{self-heuristic\xspace}

\newcommand{\vali}{probe-test\xspace}
\newcommand{\test}{test\xspace}

\newcommand{\numsample}{70,346\xspace}
\newcommand{\numtoken}{19,355,711\xspace}

\newcommand{\role}{role}
\newcommand{\reinforce}{reinforce}
\newcommand{\confirmation}{task confirmation}
\newcommand{\pos}{positive feedback}
\newcommand{\zerocot}{zero-CoT}
\newcommand{\conclusion}{right}

\newcommand{\rolestyle}[1]{\textbf{\textsf{#1}}}

\usepackage{graphicx}
\usepackage{textcomp}
\usepackage{xcolor}
\usepackage{ulem}
\usepackage{tabularx}
\usepackage{booktabs}
\usepackage{array}

\newcommand{\tblcolor}{\color{black}}

\usepackage{listings}
\definecolor{codegreen}{rgb}{0,0.6,0}
\definecolor{codedarkgreen}{rgb}{0,0.5,0.0}
\definecolor{codeblue}{rgb}{0.0, 0.0, 1.0}
\definecolor{codegray}{rgb}{0.5,0.5,0.5}
\definecolor{codepurple}{rgb}{0.58,0,0.82}
\definecolor{backcolour}{rgb}{0.95,0.95,0.95}
\lstdefinestyle{mystyle}{
	backgroundcolor=\color{backcolour},   
	commentstyle=\color{codegray},
	keywordstyle=\color{codeblue},
	numberstyle=\tiny\color{codegray},
	stringstyle=\color{codepurple},
	basicstyle=\footnotesize\it,
	breakatwhitespace=false,         
	breaklines=true,                 
	captionpos=b,                    
	keepspaces=true,                 
	numbers=left,                    
	numbersep=2pt,                  
	showspaces=false,                
	showstringspaces=false,
	showtabs=false,                  
	tabsize=2,
        columns=fullflexible,
}

\definecolor{codeback}{rgb}{0.98,0.98,0.98}
\lstdefinestyle{mycodestyle}{
	backgroundcolor=\color{codeback},
	commentstyle=\color{codegreen},
	keywordstyle=\color{codeblue},
	numberstyle=\tiny\color{codegray},
	stringstyle=\color{codepurple},
	basicstyle=\footnotesize\scriptsize,
	breakatwhitespace=false,         
	breaklines=true,                 
	captionpos=b,                    
	keepspaces=true,                 
	numbers=left,                    
	numbersep=5pt,                  
	showspaces=false,                
	showstringspaces=false,
	showtabs=false,                  
	tabsize=2,
    frame=none,
}
\definecolor{oneshotcolor}{HTML}{E6A400}
\definecolor{role}{HTML}{F57328}
\definecolor{encourage}{HTML}{00CA0D}
\definecolor{CoT}{HTML}{009DFF}
\definecolor{mock}{HTML}{AC00C3}
\definecolor{infocolor}{HTML}{F6378F}
\lstdefinestyle{mysamplestyle}{
	backgroundcolor=\color{codeback},
	commentstyle=\color{codegreen},
	keywordstyle=\color{codeblue},
	numberstyle=\tiny\color{codegray},
	stringstyle=\color{codepurple},
	basicstyle=\footnotesize\it,
	breakatwhitespace=false,         
	breaklines=true,                 
	captionpos=b,                    
	keepspaces=true,                 
    numbers=left,                    
	numbersep=5pt, 
    frame=single,
    frameround=tttt,                 
	showspaces=false,                
	showstringspaces=false,
	showtabs=false,                  
	tabsize=2,
    columns=fullflexible,
}
\usepackage{url}

\usepackage{tikz}

\newcommand{\rcolor}{\rowcolor{gray!5 }}

\usepackage{float}

\begin{document}


\date{}

\title{Exploring ChatGPT's Capabilities on Vulnerability Management\vspace{-1em}
}
\ifdefined\DRAFT
 \pagestyle{fancyplain}
 \lhead{Rev.~\therev}
 \rhead{\thedate}
 \cfoot{\thepage\ of \pageref{LastPage}}
\fi

\pagestyle{empty}


\newcommand{\equalcont}{*}
\newcommand{\corresponding}{\dag}

\author[$1,2,$\equalcont]{Peiyu Liu}
\author[$1,2,$\equalcont]{Junming Liu}
\author[$3,$\corresponding]{Lirong Fu}
\author[$4$]{Kangjie Lu}
\author[$1,2$]{Yifan Xia}
\author[$1,5$]{Xuhong Zhang}
\author[$1$]{\\Wenzhi Chen}
\author[$6$]{Haiqin Weng}
\author[$1$]{Shouling Ji}
\author[$1,2,$\corresponding]{Wenhai Wang}

\affil[ ]{\small $^1${Zhejiang University},
$^2${Zhejiang University NGICS Platform}, 
$^3${Hangzhou Dianzi University}, 
$^4${University of Minnesota},\authorcr
$^5${Jianghuai Advance Technology Center},
$^6${Ant Group}
}

\affil[ ]{{\small ~\{liupeiyu, jmliu, fulirong007\}@zju.edu.cn,~~kjlu@umn.edu,~~\{yfxia, zhangxuhong, chenwz\}@zju.edu.cn,
~~haiqin.wenghaiqin@antgroup.com,~\{sji, zdzzlab\}@zju.edu.cn}}
\vspace{-5em}
\maketitle

\newcommand\blfootnote[1]{%
	\begingroup
	\renewcommand\thefootnote{}\footnote{#1}%
	\addtocounter{footnote}{-1}%
	\endgroup
}

\setlength{\skip\footins}{5pt}
\blfootnote{\textsuperscript{\equalcont}Peiyu Liu and Junming Liu contributed equally.}
\blfootnote{\textsuperscript{\corresponding}Lirong Fu and Wenhai Wang are co-corresponding authors.}
\vspace{-2em}
\begin{abstract}
Recently, ChatGPT has attracted great attention from the code analysis domain. Prior works show that ChatGPT has the capabilities of processing foundational code analysis tasks, such as abstract syntax tree generation, which indicates the potential of using ChatGPT to comprehend code syntax and static behaviors. 
However, it is unclear whether ChatGPT can complete more complicated real-world vulnerability management tasks, such as the prediction of security relevance and patch correctness, which require an all-encompassing understanding of various aspects, including code syntax, program semantics, and related manual comments. 

In this paper, we explore ChatGPT's capabilities on 6 tasks involving the complete vulnerability management process with a large-scale dataset containing \numsample samples. For each task, we compare ChatGPT against SOTA approaches, investigate the impact of different prompts, and explore the difficulties. 
The results suggest promising potential in leveraging ChatGPT to assist vulnerability management. One notable example is ChatGPT's proficiency in tasks like generating titles for software bug reports.
Furthermore, our findings reveal the difficulties encountered by ChatGPT and shed light on promising future directions. 
For instance, directly providing random demonstration examples in the prompt cannot consistently guarantee good performance in vulnerability management. 
By contrast, leveraging ChatGPT in a self-heuristic way---extracting expertise from demonstration examples itself and integrating the extracted expertise in the prompt is a promising research direction. 
Besides, ChatGPT may misunderstand and misuse the information in the prompt. Consequently, effectively guiding ChatGPT to focus on helpful
information rather than the irrelevant content is still an open problem. 
\end{abstract}




\section{Introduction}
\label{s:intro}

Recently, there has been a notable proliferation of powerful large language models (LLMs)  owing to the rapid development of AI techniques~\cite{vaswani2017attention,Codex}.
Numerous world-leading companies such as OpenAI, Facebook, Microsoft, and many open-source maintainers have contributed significantly to the development of a large number of LLMs. 
To date, LLMs have achieved considerable success and have been widely used in diverse domains, including data augmentation~\cite{anaby2020not}, code summarization~\cite{ye2023cp}, and medical assiatant~\cite{moor2023foundation}.
Among all the various proposed LLMs, ChatGPT~\cite{ChatGPT}, an intelligent human-machine dialogue LLM, has attracted considerable attention. Notably, ChatGPT has achieved a remarkable feat by becoming the fastest-growing app globally, amassing 100 million users within two months of its launch~\cite{UBS}. 

Existing research works have demonstrated that ChatGPT exhibits outstanding performance in traditional Natural Language Processing (NLP)  tasks, including machine translation~\cite{jiao2023chatgpt}, question-answering~\cite{tan2023evaluation}, and text summarization~\cite{yang2023exploring}.
Moreover, since programming language, to some extent, shares many analogous characteristics with natural language---both of them are organized through specific grammatical structures and can express certain semantics, researchers turn to utilize ChatGPT for code-related analysis~\cite{ma2023scope,ye2024uncovering}.
For instance, Tian et al. explored ChatGPT's capabilities for code generation, program repair, and code summarization~\cite{tian2023chatgpt}. 
Additionally, Xia et al. proposed to leverage ChatGPT for automated program repair~\cite{xia2023automated}. 
Moreover, a recent study conducted by Ma et al. investigates ChatGPT's capabilities for understanding program syntax, static behaviors, and dynamic behaviors~\cite{ma2023scope}.

Despite the growing body of literature that applies ChatGPT for general software engineering, its adoption in the security domain remains underexplored. 
One of the most important areas in the security domain is vulnerability management \cite{vm}.
Current ChatGPT-focused research merely focuses on several specific tasks within vulnerability management, such as vulnerability fixing~\cite{pearce2021examining}.
However, vulnerability management encompasses a comprehensive lifecycle that consists of complex phases, each presenting its unique set of challenges~\cite{shahzad2012large}. 
Without a holistic horizon of the whole lifecycle, the capacity of ChatGPT to bolster this vital security process remains an unknown question. 

To fill this research gap, in this paper, we explore: \textbf{Can ChatGPT directly assist software maintainers in diverse tasks during the whole vulnerability management process?} We focus on vulnerability management as it is important, presents complexities, and requires significant manual efforts~\cite{mozilla}. 
Given the similarities between code understanding tasks in vulnerability management and NLP, ChatGPT's application in solving these tasks is highly plausible.
Concretely, we want to investigate if ChatGPT can achieve capability on par with the state-of-the-art (SOTA) approaches for vulnerability management tasks. 
Besides, considering the impact of existing prompt engineering methods, we also aim to systematically investigate their effect on ChatGPT's performance. 
Finally, for the difficulties encountered by ChatGPT for vulnerability management, we aim to shed light on aspects for future exploration. 
In summary, this paper measures ChatGPT's performance on various vulnerability management tasks from the following three perspectives. 
\textbf{RQ1}: Does ChatGPT achieve capability on par with the SOTAs?
\textbf{RQ2}: How do prompt engineering methods impact ChatGPT's performance?
\textbf{RQ3}: What is the promising future direction to improve ChatGPT's performance on each task?

To answer these research questions, we compared ChatGPT's performance to 11 SOTA approaches on 6 vulnerability management tasks, including bug report summarization, security bug report identification, vulnerability severity evaluation, vulnerability repair, patch correctness assessment, and stable patch classification. 
To achieve this, we first collect the dataset provided by the SOTA approaches of each task. The dataset used in this paper contains \numsample samples, \numtoken tokens in total. By leveraging this dataset, we evaluate ChatGPT's performance for each task with the same metrics used by each SOTA approach. Then, we investigate the influence of different prompt engineering methods. Finally, we analyze ChatGPT's responses to identify the bottlenecks of each task. 

Our evaluation and analysis results demonstrate that (1) ChatGPT can outperform the SOTA approaches without being trained specifically for some vulnerability management tasks, especially for tasks related to software document processing, e.g., summarizing bug reports. (2) ChatGPT can achieve comparable capabilities to the SOTA approaches with the help of prompts that contain manual knowledge for some tasks, e.g., security bug report identification. (3) Providing random demonstration examples in the prompt oftentimes achieves limited performance. By contrast, leveraging ChatGPT in a self-heuristic way---extracting expertise from demonstration examples itself and integrating the extracted expertise in the prompt is a promising research direction on some tasks, e.g., vulnerability severity evaluation. 
(4) Intuitively, the more information provided in the prompt, the better ChatGPT performs. However, our investigation reveals that providing excessive information can lead to misunderstandings and misuse by ChatGPT. Therefore, directing ChatGPT to prioritize relevant and constructive information over potentially problematic content is a critical area for further research. Our contributions are as follows. 

(1) We conduct the first large-scale evaluation of ChatGPT for vulnerability management tasks. The results indicate the desirable prospects of leveraging ChatGPT to assist vulnerability management. 

(2) We investigate the influence of various prompt engineering methods for different vulnerability management tasks, which can provide helpful suggestions on designing better prompts to exploit ChatGPT for each task thoroughly. 

(3) We uncover the bottlenecks encountered by ChatGPT on vulnerability management and shed light on promising future directions to improve ChatGPT's performance.

\section{Background}
\label{s:background}

\begin{figure*}[t]
    \centering
    \includegraphics[width=0.98\linewidth]{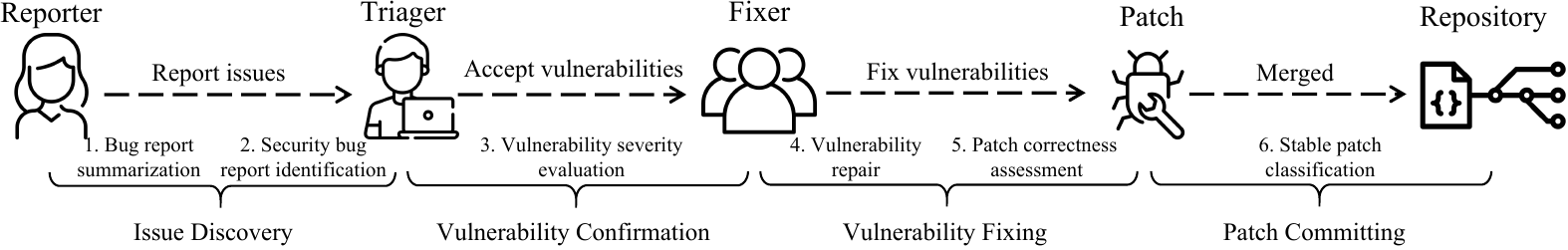} \vspace{-8pt}
    \caption{The vulnerability management process. }
    \label{fig:lifecycle}
\end{figure*}

\subsection{Vulnerability Management Process}
\label{sub-s:defect}

Vulnerability management constitutes a crucial process in software development, encompassing identifying, classifying, and mitigating vulnerabilities in software products~\cite{vm,mozilla}. 
As shown in~\autoref{fig:lifecycle}, a well-established vulnerability management process adopted by prominent software development teams, such as Mozilla \cite{bugzilla}, involves at least four general phases: issue discovery, vulnerability confirmation, vulnerability fixing, and patch committing.

\textbf{Issue Discovery.}
In this phase, an issue \textit{reporter} reports issues through the bug tracking system or the version control system such as Github \cite{github}. 
Subsequently, a \textit{triager} examines the reported issues for detailed assessment.
In real-world scenarios, the \textit{triager} may encounter a substantial influx of bug reports. 
The thorough evaluation of the entire report is quite time-consuming. 
Hence, succinct and accurate \textbf{bug report summarization} is pivotal for the \textit{triager} to swiftly grasp the essence of the bug~\cite{chen2020stay}. 
Unfortunately, there usually exist quality concerns of bug report summaries submitted by the \textit{reporters} due to their various professionalism and comprehension of the projects. 

Besides, since the vulnerability management process focuses on security issues, efficient \textbf{security bug report identification} among an overwhelming number of bug reports is another crucial task for the \textit{triager}. 
Nevertheless, distinguishing security-related bug reports requires in-depth domain-specific knowledge and considerable human effort, making it a strong demand for automated identification~\cite{wu2021data,zheng2022domain}. 


\textbf{Vulnerability Confirmation.}
During the vulnerability confirmation phase, a \textit{triager}, usually a senior developer, is responsible for preliminarily confirming the existence of the reported vulnerabilities. 
Afterward, the \textit{triager} will assign the bug-fixing task to an appropriate \textit{fixer} in order of severity. 

To prevent potential exploitation, all confirmed reports should be assigned timely.
However, when confronted with a large number of vulnerabilities, it is impractical to patch all of them within a limited time and workforce. 
Given the limited resources, both the \textit{triager} and the \textit{fixer} need to prioritize more severe vulnerabilities, as they pose higher security risks to the software. 
Consequently, \textbf{vulnerability severity evaluation} becomes the primary step in handling them~\cite{wu2022aware}. 


\textbf{Vulnerability Fixing.}
In this phase, a vulnerability \textit{fixer} generates patches to repair the assigned vulnerabilities. 
However, patch development demands a profound comprehension of the code context and underlying logic, making it challenging. 
To support the \textit{fixers} in efficiently fixing software errors, automated \textbf{vulnerability repair} tools become crucial as they empower maintainers to produce high-quality patches efficiently~\cite{pearce2021examining}.
Moreover, the process extends beyond mere repair, considering 
the prevalence of incorrect patches that fail to address vulnerabilities adequately or introduce new complications~\cite{tian2022change}. 
Therefore, \textbf{patch correctness assessment} emerges as a critical step in this phase~\cite{le2023invalidator}.


\textbf{Patch Committing.}
Besides the confirmed vulnerability patches generated from the previous process, software maintainers may receive patches from third-party developers, which need to be classified and applied to the suitable codebase for certain users~\cite{hoang2021patchnet}.  
Typically, software patches can be categorized into stable (bug-fixing) patches and feature enhancement patches. 
For instance, the Linux kernel maintains a series of stable versions that accept only stable patches.
As patches for stable versions contain fixes for bugs that can impact security and stability, it is important to ensure the correctness of \textbf{stable patch classification}. However, facing a large number of patches, identifying stable patches accurately and efficiently becomes a significant challenge in this phase. 


\textbf{Summary.}
Vulnerability management comprises several phases, each presenting its unique set of challenges. Within the scope of these phases, we have pinpointed six pivotal tasks that epitomize the core difficulties encountered in this domain, as listed in \autoref{tab:baseline}.
Meanwhile, developing automatic tools based on traditional software analysis techniques and machine learning models faces significant challenges since these tasks require an in-depth understanding of complex code semantics, program logic, software documents, etc. 
Considering ChatGPT's established capabilities in code generation and interpretation, it is imperative to investigate its performance across these vulnerability management tasks. Such an investigation could unveil new avenues for employing ChatGPT in this domain and catalyze further research.

\begin{table}
    \small
	\centering
	\caption{Tasks, baselines, and dataset. S = Sample. T = Token.}
	\label{tab:baseline}
	\begin{tabular}{cccc}
		\toprule
         \multirow{2}{*}{\textbf{Task}}  & \multirow{2}{*}{\textbf{Baseline}} & \multicolumn{2}{c}{\textbf{ Dataset}} \\
        \cmidrule(lr){3-4}
        &&\textbf{\# S} & \textbf{\# T} \\
        \midrule
\rcolor          \colorbox{gray!5 }{\makecell[c]{Bug report\\summarization}} & iTAPE~\cite{chen2020stay} & 33,438 & 6,176,326 \\
          & Farsec~\cite{wu2021data} & & \\
          & DKG~\cite{zheng2022domain} & 22,970 & 5,686,564\\
           \multirow{-3}{*}{\makecell[c]{Security bug\\report identification}} & CASMS~\cite{ma2022casms} & & \\
\rcolor                 \colorbox{gray!5 }{\makecell[c]{Vulnerability\\severity evaluation}} & DiffCVSS~\cite{wu2022aware} & 1,642 & 82,397 \\
          & LLMset~\cite{pearce2021examining} & & \\
          \multirow{-2}{*}{\makecell[c]{Vulnerability\\repair}} & ExtractFix~\cite{gao2021beyond} &  \multirow{-2}{*}{12} & \multirow{-2}{*}{10,601} \\
\rcolor                 & Quatrain~\cite{tian2022change}  & 995 & 468,739 \\
\rcolor                 & Invalidator~\cite{le2023invalidator} &  139 & 31,663 \\
\rcolor                 \multirow{-3}{*}{\makecell[c]{Patch correctness\\assessment}} & Panther~\cite{tian2023best} &  208 & 45,204 \\
                  \makecell[c]{Stable patch\\classification} & PatchNet~\cite{hoang2021patchnet} & 10,896 & 6,854,217 \\
        \midrule
        \rcolor \textbf{Total}&\textbf{11}  & \textbf{\numsample} & \textbf{\numtoken} \\
	\bottomrule
	\end{tabular}
\end{table}

\subsection{ChatGPT and Prompt}
\label{sub-s:chatgpt}

\begin{figure*}
    \centering
    \includegraphics[width=0.90\textwidth]{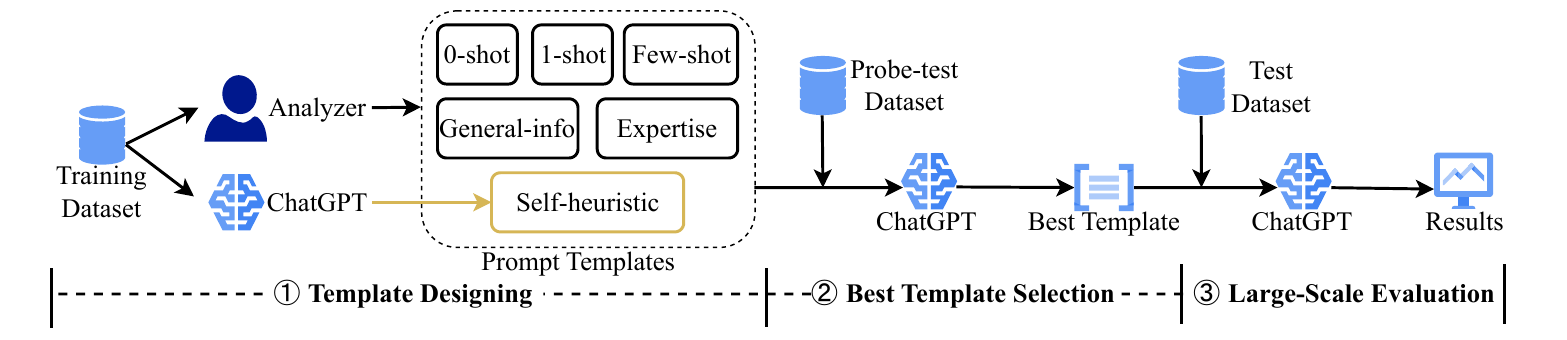}
    \caption{Evaluation pipeline.}
    \label{fig:workflow}
\end{figure*}

ChatGPT is an artificial intelligence chatbot trained to provide human-like responses to users' questions in a conversational way. 
Users can use ChatGPT through its web interface~\cite{ChatGPT} or official API~\cite{API}. 
A common strategy for applying general models to specific tasks is model fine-tuning. 
However, this approach is often eschewed due to its labor-intensive nature and significant resource consumption~\cite{finetuning}.
Consequently, attention has shifted toward optimizing the prompt, i.e., the input of ChatGPT, which significantly influences the relevance and accuracy of the ChatGPT's output~\cite{Prompt}.

Currently, there are a series of prompt construction strategies employed for enhancing the capability of ChatGPT.
Among these, in-context learning has emerged as the dominant paradigm~\cite{icl2023survey}. 
The foundational approach to in-context learning, known as \textbf{\zeroshot} prompting, instructs ChatGPT by directly describing the task and the associated question~\cite{pearce2021examining}. 
While \zeroshot prompting has shown promising performances by leveraging the prior knowledge from the training resource, it sometimes falters when confronted with unfamiliar tasks. 

To address this problem, researchers devised advanced prompts by integrating demonstrations, allowing ChatGPT to discern patterns inherent to specific tasks~\cite{icl2023survey}.
Depending on the volume of demonstration examples within a prompt, they can be categorized into \textbf{\oneshot} prompting (with a singular demonstration example) or \textbf{\fewshot} prompting (incorporating multiple demonstration examples)~\cite{gao2023makes}.
While well-organized demonstrations have proven effective for simple tasks, they tend to be less efficient for intricate tasks demanding complex logic and domain expertise.
To counter these challenges, another line of works enhance ChatGPT by utilizing refined demonstration formatting.
This includes the provision of supplemental \textbf{general information}, such as role definitions \cite{pearce2021examining,xia2023automated}, and integrating \textbf{domain-specific expertise}, such as vulnerability patterns \cite{sun2023gpt}.

In addition to the prompt, the foundational model of ChatGPT also impacts its performance. Users can choose to use \gptthree or \gptfour when using ChatGPT~\cite{API}. 
A systematic investigation of the effect of different prompts and models for vulnerability management tasks is still lacking. Exploring this problem is essential for thoroughly exploiting ChatGPT for vulnerability management. 
\section{Evaluation Framework}
\label{s:framework}

\subsection{Evaluation Pipeline}

\autoref{fig:workflow} shows the pipeline of our evaluation, which includes three phases: \textcircled{1} template design, \textcircled{2} best template selection, and \textcircled{3} large-scale evaluation.  

Currently, automatic prompt generation~\cite{shin2020autoprompt} is an ongoing research work that has not been well addressed. Consequently, in phase \textcircled{1}, for each evaluated task, according to the construction rules outlined in~\autoref{sub-s:chatgpt}, we first design the prompt templates listed in~\autoref{tab:templates} manually based on the heuristics derived from existing widely adopted strategies~\cite{promptexample,gao2023makes}. 
To ensure the effectiveness of each prompt template, we assess them using 100 random samples from the training dataset. Subsequently, we refine the templates based on our manual analysis of ChatGPT's responses. Details regarding the template implementation process are discussed in~\autoref{sub-s:prompt}. 
As a result, we acquire a collection of prompt templates that demonstrate consistent and reliable performance on the training dataset.

In phase \textcircled{2}, we evaluate all prompt templates with the \vali dataset to select the best prompt template. To ensure a representative subset of the entire dataset, we construct a \vali dataset by separating other 10\% random samples from the training dataset. However, for certain tasks, the extensive nature of the training dataset used by the SOTA approaches presents challenges. Therefore, for these tasks, considering the time and cost constraints, we limit the \vali dataset to 1,000 samples. Besides, for vulnerability repairing task, since there is no training dataset used by the SOTA approaches, we opt to construct the \vali dataset using hand-crafted vulnerabilities provided by ~\cite{pearce2021examining}.
Finally, in phase \textcircled{3}, to fully explore ChatGPT's potential, we use the prompt template that yields the best performance in the probe-test to conduct a large-scale evaluation on the test dataset.\footnote{Our evaluations are conducted with ChatGPT official API~\cite{API}. Specifically, We use ChatGPT based on gpt-3.5-turbo-0301 in phases \textcircled{1} and \textcircled{2} since it is more economical than gpt-4. The large-scale tests are conducted with ChatGPT based on gpt-4-0314. We set temperature = 0 and top\_p = 1.0 to enhance determinism.}

\begin{table*}[ht]
    \small 
    \centering
    \caption{Templates for task prompt generation.} 
    \label{tab:templates}
    \setlength{\tabcolsep}{0.2cm}
    \begin{tabularx}{\textwidth}{>{\centering}m{1.8cm}>{\raggedright}m{6.5cm}>{\raggedright}m{\textwidth-9.5cm}}
        \toprule
        \textbf{Template Name} & \textbf{Description} & \textbf{Template} \tabularnewline
        \midrule
        \zeroshot & Describes the task and directly gives the input query. &  \rolestyle{USER} <task description> <input>   \tabularnewline
        \midrule
        \oneshot & Describes the task and provides a random-selected demonstration example before query. &  \rolestyle{USER} <task description> <demonstration example>   <input>    \tabularnewline
        \midrule
        \fewshot & Describes the task and provides multiple random-selected demonstration examples before query. &  \rolestyle{USER} <task description> <demonstration example 1> <demonstration example 2> <demonstration example 3> <demonstration example 4>  <input>    \tabularnewline
        \midrule
        \midrule
        \prompteng & Integrates the task-related role assignment into task description through system instruction and complements zero-CoT instructions before query.   &   \rolestyle{SYSTEM} <\role> <task description> <\reinforce>\\ \rolestyle{USER} <task description> <\confirmation>\\ \rolestyle{ASSYSTANT} <\confirmation>\\ \rolestyle{USER} <\pos> <input> <\zerocot> <\conclusion> \tabularnewline
        \midrule
        \manualinfo & Based on the \prompteng template, provides manually summarized domain-specific expertise with task description.   &   \rolestyle{SYSTEM} <\role> <task description> <expertise> <\reinforce>\\ \rolestyle{USER} <expertise> <task description> <\confirmation>\\ \rolestyle{ASSYSTANT} <\confirmation>\\ \rolestyle{USER} <\pos> <input> <\zerocot> <\conclusion> \tabularnewline
        \midrule
        \midrule
        \gptinfo &  Integrates ChatGPT-summarized domain-specific knowledge into task description before query.  &   \rolestyle{SYSTEM} <\role> <task description> <\reinforce>\\ \rolestyle{USER} <knowledge> <task description> <\confirmation>\\ \rolestyle{ASSYSTANT} <\confirmation>\\ \rolestyle{USER} <\pos> <input> <\zerocot> <\conclusion> \tabularnewline
        \bottomrule
    \end{tabularx}
\end{table*}

\subsection{SOTA Approaches and Dataset}

To assess ChatGPT's capabilities on the six vulnerability management tasks described in~\autoref{sub-s:defect}, we collect the SOTA approaches for each task as the baselines. The SOTA approaches are derived from the top venues in the fields of security, programming language, and machine learning over the past three years. By sourcing approaches from these reputable venues, we ensure that our evaluation encompasses the most current and relevant advancements in the respective domains, facilitating a comprehensive and up-to-date analysis of ChatGPT's capabilities on vulnerability management. As shown in \autoref{tab:baseline} we include 11 SOTA approaches for the six evaluated tasks. Then, we evaluate ChatGPT with the same metrics used by each SOTA approach to conduct the comparison. 

To ensure a fair comparison between ChatGPT and SOTA approaches, we obtain the same dataset used in each corresponding SOTA paper for each task. Specifically, we strictly separate the training dataset and test dataset according to the original setting in those papers. 
Totally, the test dataset used in this paper contains \numsample samples (\numtoken tokens).  

\subsection{Prompt Design and Implementation}
\label{sub-s:prompt}
In~\autoref{tab:templates}, we provide detailed explanations on the methodology employed for constructing prompt templates.
For \oneshot and \fewshot prompts, the demonstration samples are randomly selected from the training dataset. Particularly, according to~\cite{gao2023makes}, we select four demonstration samples for \fewshot templates to obtain considerably good performance. 
In the general-information template, we include the instructions that show their superiority in traditional NLP tasks~\cite{Prompt}, such as zero-CoT~\cite{kojima2022large} and ChatGPT's role \cite{Prompt} in the prompt (\autoref{tab:skill} lists the instructing skills used in the \prompteng template, which is deferred to~\autoref{a:a}).
As illustrated in \autoref{lst:manualinfo}, the prompt leverages many existing methods, such as the role assignment in line 1.
In the \manualinfo prompt template, we provide domain expert knowledge in the prompt. The offered expertise is obtained from the related documentations~\cite{CVSS,linux} and literatures~\cite{wu2021data,chen2020stay} (\autoref{tab:expertise} provides the expertise content for each task, which is deferred to~\autoref{a:a}). As shown in \autoref{lst:manualinfo}, the \manualinfo prompt differs from the \prompteng prompt by providing the characteristic of security bug reports (lines 3 - 6 and 9 - 12). In this example, we guide ChatGPT that memory leak should be treated as a security bug.

It is worth noting that for some evaluated tasks, summarizing expert knowledge can be non-trivial. For example, it is hard to summarize the rules of determining vulnerability severity. In such situations, we turn to explore the potential of guiding ChatGPT by leveraging the knowledge summarized by itself. Thus, we propose the \gptinfo prompt template. Specifically, we provide several demonstration examples to ChatGPT and ask it to summarize knowledge from these examples. 
For example, we provide 100 labeled function descriptions to ChatGPT and tell it to \textit{``summarize the characteristics of the function descriptions that should map to the CVSS AV:Network metric''}. Then, we can obtain knowledge summarized by ChatGPT, like 
``Functions that involve network communication, socket handling, or network
device management. Examples: sock\_register, udp4\_hwcsum, ...'' (\autoref{fig:selfprompt} provides an illustrative example that elucidates the extraction process, which is deferred to~\autoref{a:a}). 
After that, we can provide the summary in the \gptinfo prompt in the same way as the \manualinfo prompt. 
More comprehensive descriptions, including specific examples for each prompt template, are available in~\autoref{tab:templateexample2} (deferred to~\autoref{a:a}).
We will provide all the prompts on \gitrepo to support further research. 

\section{Evaluation Results}

In this section, we elaborate on the evaluation results of ChatGPT for the evaluated tasks. For each task, we seek the answers to the research questions proposed in~\autoref{s:intro} by comparing the performance of ChatGPT and SOTA approaches (\textbf{RQ1}), investigating the impact of different prompt templates (\textbf{RQ2}), and exploring the potential future research directions to address the bottlenecks encountered by ChatGPT (\textbf{RQ3}). 

\subsection{Bug Report Summarization}
\label{sub-s:title-generation-result}

\begin{table*}    
    \small
	\centering
    \tblcolor
	\caption{The evaluation result on bug report summarization. }
	\label{tab:title-generation}
	\setlength{\tabcolsep}{1.9mm}     
 	{
	\begin{tabular}{ccccccccccccccccccccc}
		\toprule
        {\multirow{2}{*}{\textbf{Approach}}} & {\multirow{2}{*}{\textbf{Prompt}}} & {\multirow{2}{*}{\textbf{Dataset}}} &  \multicolumn{3}{c}{\textbf{ROUGE-1}} & \multicolumn{3}{c}{\textbf{ROUGE-2}} & \multicolumn{3}{c}{\textbf{ROUGE-L}} \\
        \cmidrule(lr){4-6} \cmidrule(lr){7-9} \cmidrule(lr){10-12}
        \multicolumn{3}{c}{} & F1 & Precision & Recall & F1 & Precision & Recall & F1 & Precision & Recall
        \\
        \midrule
        \rcolor 
        {iTAPE~\cite{chen2020stay}} & - & \test & 31.36 & 32.61 & 31.72 & 13.12 & 13.77 & 13.34 & 27.79 & 30.10 & 29.32 \\
        \midrule 
        \gptthree & \zeroshot & \vali & 34.33 & 30.54 & 42.11 & 11.05 & 9.66 & 13.99 & 27.95 & 24.78 & 34.41 \\ 
        \rcolor \gptthree & \oneshot & \vali & 36.82 & 33.54 & 43.67 & 13.27 & 11.97 & 16.13 & 30.86 & 28.03 & 35.71 \\ 
        \gptthree & \fewshot & \vali & 37.30 & 33.91 & 44.26 & 13.99 & 12.61 & 16.92 & 31.52 & 28.57 & 37.53 \\ 
        \rcolor \gptthree & \prompteng & \vali & 32.37 & 28.23 & 41.12 & 10.73 & 9.25 & 14.10 & 26.55 & 23.10 & 33.83 \\   
        \gptthree & \manualinfo & \vali & 33.27 & 29.50 & 41.23 & 11.30 & 9.87 & 14.37 & 27.58 & 24.35 & 34.32 \\ 
        \rcolor \gptthree & \gptinfo & \vali & 33.08 & 30.25 & 40.16 & 11.26 & 10.28 & 13.88 & 27.53 & 25.10 & 33.56 \\ 
        \gptfour & \fewshot & \vali & \textbf{40.38} & \textbf{39.07} & \textbf{44.35} & \textbf{15.86} & \textbf{15.26} & \textbf{17.69} & \textbf{34.30} & \textbf{33.12} & \textbf{37.75} \\ 
    \midrule
        \rcolor \gptfour & \fewshot & \test & 39.17 & 37.52 & 43.45 & 14.34 & 13.58 & 16.35 & 33.23 & 31.77 & 36.92 \\
	\bottomrule
	\end{tabular}}
\end{table*}

In this evaluation, we ask ChatGPT to summary a given bug report in each query.

\begin{figure}
\begin{lstlisting}[style=mysamplestyle, caption={},label={}, escapeinside=@@]
@{\footnotesize{\ttfamily{\textbf{SYSTEM}}}\qquad\textcolor{role}{You are Frederick, an AI expert in bug report analysis.}}@ Your
@{\qquad\qquad\quad}@task is to decide whether a given bug report is a security bug 
@{\qquad\qquad\quad}@report (SBR) or non-security bug report (NBR).  @{\textcolor{infocolor}{\textbf{When}}}@
@{\qquad\qquad\quad}@@{\textcolor{infocolor}{\textbf{analyzing the bug report, take into account that bug reports}}}@
@{\qquad\qquad\quad}@@{\textcolor{infocolor}{\textbf{related to memory leak or null pointer problems should be}}}@
@{\qquad\qquad\quad}@@{\textcolor{infocolor}{\textbf{seen as security bug report.}}}@ @{\textcolor{encourage}{Remember, you're the best AI bug}}@ 
@{\qquad\qquad\quad}@@{\textcolor{encourage}{report analyst and will use your expertise to provide the best}@
@{\qquad\qquad\quad}@@{\textcolor{encourage}{possible analysis.}@
@{\footnotesize{\ttfamily{\textbf{USER}}}\quad\qquad}@@{\textcolor{infocolor}{\textbf{A security bug report is a bug report describing one or more}}}@
@{\qquad\qquad\quad}@@{\textcolor{infocolor}{\textbf{vulnerabilities of a software. Besides, bug reports that directly}}}@
@{\qquad\qquad\quad}@@{\textcolor{infocolor}{\textbf{mention "memory leak" or "null pointer" problems must be}}}@
@{\qquad\qquad\quad}@@{\textcolor{infocolor}{\textbf{seen as security bug reports.}}}@ I will  give you a bug report and 
@{\qquad\qquad\quad}@you will analyze it, step-by-step, to know whether or not it is 
@{\qquad\qquad\quad}@a security bug report. @{\textcolor{mock}{Got it?}}@
@{\footnotesize{\ttfamily{\textbf{ASSISTANT}}}}@  @{\textcolor{mock}{Yes, I understand. I am Frederick, and I will analyze the bug}}@
@{\qquad\qquad\quad}@ @{\textcolor{mock}{report.}}@
@{\footnotesize{\ttfamily{\textbf{USER}}}\quad\qquad}@@{\textcolor{encourage}{Great! Let's begin then :)}}@
@{\qquad\qquad\quad}@For the bug report:
@{\qquad\qquad\quad}@<bug report>
@{\qquad\qquad\quad}@---------
@{\qquad\qquad\quad}@Is this bug report (A) a security bug report (SBR), or (B) a
@{\qquad\qquad\quad}@non-security bug report (NBR).
@{\qquad\qquad\quad}@Answer: Let's think @{\textcolor{CoT}{step-by-step}}@ to @{\textcolor{encourage}{reach the right conclusion}}@, 
\end{lstlisting}\vspace{-8pt}
\caption{An example of the \manualinfo prompt. After removing the \textcolor{infocolor}{\textbf{bold pink text}}, the rest represents the \prompteng prompt.}
\label{lst:manualinfo}\vspace{-8pt}
\end{figure}

\textbf{ChatGPT's Performance.} \autoref{tab:title-generation} reports the results of this evaluation, which demonstrate that ChatGPT can achieve outstanding performance in this task. 
For instance, the recall score of ChatGPT based on \gptthree with the \zeroshot prompt on the \vali dataset is 42.11, 13.99, and 34.41 under ROUGE-1, ROUGE-2, and ROUGE-L, respectively, which is better than that of the SOTA approach (31.72, 13.34, 29.32). Moreover, its F1 scores under ROUGE-1 (34.33) and ROUGE-L (27.95) also outperform the SOTA approach (31.36 and 27.79).
The results indicate that ChatGPT can generate high-quantity summaries for bug reports even with the most straightforward prompt (\zeroshot). This is reasonable since bug report summarization is similar to traditional NLP tasks, such as summarizing news
, on which ChatGPT and other LLMs have shown excellent performance~\cite{zhang2020pegasus}. 
Hence, the results encourage software maintainers to leverage ChatGPT for summarizing bug report and other vulnerability management tasks related to natural language processing. 

\textbf{The Impact of Prompts and Models.} From~\autoref{tab:title-generation}, we observe that the most straightforward prompt (\zeroshot) already enables ChatGPT to achieve outstanding performance, whereas more complex prompts do not consistently enhance ChatGPT's performance.
Overall, the \fewshot prompt templates perform better than other templates in this task. 
For example, for ChatGPT based on \gptthree evaluated on the \vali dataset, its ROUGE-L F1 score with the \zeroshot prompt is 34.33. This score improves to 37.30 with the \fewshot prompt and decreases to 32.37 with the \prompteng prompt.
This observation indicates that providing demonstration examples is helpful for ChatGPT to generate high-quality bug report summarization. 

In addition to the prompt evaluation, we also explore the impact of different foundational models. As shown in~\autoref{tab:title-generation}, when using the \fewshot prompt on the \vali dataset, the F1 scores of ChatGPT based on \gptfour are 40.38, 15.86, and 34.30, which are better than that of ChatGPT based on \gptthree (33.08, 11.26, and 27.53).
The results encourage us to use \gptfour when it is available.
Moreover, it is important to recognize that reliance solely on automated metrics such as ROUGE does not necessarily guarantee the quality of summarizations from a human reader's perspective. To address this issue, we have also implemented a user study to verify our findings, which is detailed in Section \ref{subsec:userstudy}.

\textbf{Implications.} In this evaluation, when using the prompt templates that refine demonstration formatting (\prompteng, \manualinfo, and \gptinfo), most metric scores are even worse than the \zeroshot template. 
This experimental result demonstrates that the original training task of ChatGPT has already equipped itself with the ability to comprehend this summarization task. Therefore, ChatGPT naturally excels at comprehending, generating, and summarizing text. In this context, providing a straightforward task description and a relevant question can yield strong performance. Conversely, \textit{introducing additional information and expertise in the prompt can lead to confusion for ChatGPT in understanding this task.}

\begin{table}
    \small
    \tblcolor
	\centering
	\caption{The evaluation result on security bug report identification. R = Recall. P = Precision. FPR = False Positive Rate. G = G-measure. }
	\label{tab:security-bug-report}
	\setlength{\tabcolsep}{0.7mm} 
 	{
	\begin{tabular}{cccccccccccccccccccccc}
		\toprule
        \textbf{Approach} & \textbf{Prompt} & \textbf{Dataset} & \textbf{R} & \textbf{FPR} & \textbf{P} & \textbf{F1} & \textbf{G} \\
        \midrule
        \rcolor  DKG~\cite{zheng2022domain} & - & \test & 0.70 & 0.02 & \textbf{0.74} & \textbf{0.71} & 0.81 \\
        CASMS~\cite{ma2022casms} & - & \test & 0.73 & 0.28 & - & - & 0.72 \\
        \rcolor Farsec~\cite{wu2021data} & - & \test & 0.57 & 0.16 & 0.40 & 0.43 & 0.64 \\
        \midrule
        \gptthree & \zeroshot & \vali & 0.35 & 0.02 & 0.21 & 0.27 & 0.52 \\ 
        \rcolor \gptthree & \oneshot & \vali & 0.76 & 0.09 & 0.12 & 0.21 & 0.83 \\ 
        \gptthree & \fewshot & \vali & 0.88 & 0.06 & 0.21 & 0.34 & 0.91 \\
        \gptthree & \prompteng & \vali & 0.29 & 0.01 & 0.26 & 0.28 & 0.45 \\
        \rcolor \gptthree & \manualinfo & \vali & 0.71 & 0.01 & 0.57 & 0.63 & 0.82 \\
        \gptthree & \gptinfo & \vali & 0.29 & \textbf{0.00} & 0.56 & 0.38 & 0.45 \\
        \rcolor \gptfour & \manualinfo & \vali & \textbf{0.94} & 0.04 & 0.27 & 0.42 & \textbf{0.95} \\ 
    \midrule
        \gptfour & \manualinfo & test & 0.68 & 0.04 & 0.53 & 0.57 & 0.79 \\ 
	\bottomrule
	\end{tabular}}
\end{table}

\subsection{Security Bug Report Identification}
\label{sub-s:security-prediction-result}

In this evaluation, we ask ChatGPT to answer whether a given bug report is security-related in each query. This evaluation uses three SOTA approaches, DKG~\cite{zheng2022domain}, CASMS~\cite{ma2022casms}, and Farsec~\cite{wu2021data}. The used metrics are recall, false positive rate (FPR), precision, F1, and G-measure. Among these five performance metrics, recall, precision, f1, and G-measure are the higher, the better, while FPR is the lower, the better. 

\begin{table*}
    \small
    \tblcolor
	\centering
	\caption{The evaluation result on vulnerability severity evaluation. AV = Attack Vector. AC = Attack Complexity. PR = Privileges Required. UI = User Interaction. R = Recall. P = Precision. }
	\label{tab:vulnerability-severity}
	\setlength{\tabcolsep}{0.8mm} 
 	{
	\begin{tabular}{cccccccccccccccccccccc}
		\toprule
        {\multirow{3}{*}{\textbf{Approach}}} & {\multirow{3}{*}{\textbf{Prompt}}} & {\multirow{3}{*}{\textbf{Dataset}}} & \multicolumn{6}{c}{\textbf{AV}} & \multicolumn{2}{c}{\textbf{AC}} & \multicolumn{2}{c}{\textbf{PR}}  & \multicolumn{2}{c}{\textbf{UI}} \\
        \cmidrule(lr){4-9}
        \multicolumn{3}{c}{} & \multicolumn{2}{c}{Network} & \multicolumn{2}{c}{Adjacent} & \multicolumn{2}{c}{Physical} & \multicolumn{2}{c}{High} & \multicolumn{2}{c}{High} & \multicolumn{2}{c}{Required}\\
        \cmidrule(lr){4-5} \cmidrule(lr){6-7} \cmidrule(lr){8-9} \cmidrule(lr){10-11} \cmidrule(lr){12-13} \cmidrule(lr){14-15}
        \multicolumn{3}{c}{} & R & P & R & P & R & P & R & P & R & P & R & P \\
        \midrule
        \rcolor 
        DiffCVSS~\cite{wu2022aware} & - & \test & 0.9242 & \textbf{0.9384} & 0.8750 & 0.9333 & 0.8852 & 0.9153 & \textbf{0.9151} & \textbf{0.9238} & \textbf{0.9452} & 0.9324 & \textbf{0.9167} & \textbf{0.9296} \\
    \midrule
        \gptthree & \zeroshot & \vali & 0.7143 & 0.5556 & 0 & N/A & 0 & N/A & 0.4286 & 0.6923 & 0.3684 & 0.5000 & 0 & N/A \\
        \rcolor \gptthree & \oneshot & \vali & \textbf{1.0000} & 0.2206 & 0 & N/A & 0.0909 & \textbf{1.0000} & 0.2857 & 1.0000 & 0.1053 & \textbf{1.0000} & 0.2667 & 0.3077 \\
        \gptthree & \fewshot & \vali & \textbf{1.0000} & 0.4285 & 0.4444 & 0.6667 & 0.3636 & 0.4444 & 0.6190 & 0.6842 & 0.2632 & 0.3333 & 0.6667 & 0.2703 \\
        \rcolor \gptthree & \prompteng & \vali & 0.7857 & 0.4783 & 0 & N/A & 0.1667 & 0.5000 & 0.8095 & 0.3269 & 0.7368 & 0.2188 & 0.4000 & 0.3000 \\
        \gptthree & \manualinfo & \vali & 0.8571 & 0.5714 & 0.5000 & 0.6667 & 0.0833 & \textbf{1.0000} & 0.8095 & 0.2982 & 0.5263 & 0.3704 & 0.2667 & 0.2857 \\
        \rcolor \gptthree & \gptinfo & \vali & \textbf{1.0000} & \textit{0.7368} & 0.7500 & \textbf{1.0000} & \textbf{1.0000} & 0.9231 & 0.8095 & 0.5484 & 0.8421 & 0.6400 & 0.9333 & 0.5000 \\
        \gptfour & \gptinfo & \vali & \textbf{1.0000} & \textit{0.7368} & \textbf{1.0000} & \textbf{1.0000} & 0.9167 & 0.9167 & 0.9048 & 0.6786 & 0.8947 & 0.7083 & 0.8667 & 0.7647 \\
    \midrule 
        \rcolor \gptfour & \gptinfo & \test & 0.9848 & 0.7738 & 0.9063 & 0.9355 & 0.9167 & 0.8333 & 0.7961 & 0.7321 & 0.8941 & 0.7917 & 0.7714 & 0.8852 \\
	\bottomrule
	\end{tabular}}
\end{table*}

\textbf{ChatGPT's Performance.} 
\autoref{tab:security-bug-report} shows ChatGPT's performance in this task. Generally, with advanced prompt templates and models, ChatGPT can outperform two baselines, CASMS and Farsec. However, ChatGPT cannot achieve capability on par with DKG. 
For example, the F1 and G-measure scores of ChatGPT based on \gptfour with the \manualinfo prompt on the test dataset are 32.6\% and 23.4\% higher than that of Farsec, respectively. By contrast, these scores are 19.7\% and 2.5\% lower than that of DKG, respectively.
Hence, our investigation indicates that ChatGPT has a certain, yet unsatisfactory, capability to identify security-related bug reports.

\textbf{The Impact of Prompts and Models.}
In this evaluation, the impact of prompt templates and models is complicated. First, compared to the \zeroshot prompt, some advanced prompt templates increase several metric scores while decreasing others. 
For example, for ChatGPT based on \gptthree tested on the \vali dataset, the \prompteng prompt increases the precision score while decreasing the recall score compared to the \oneshot prompt. 
We find the reason is that the \prompteng prompt makes ChatGPT very ``conservative'' for this task---ChatGPT only marks 14 tested reports as security-related with only 5 true positives in them. 
Second, the advanced model, \gptfour, also increases several metric scores while decreasing others. 
For instance, when using the same \manualinfo prompt on the \vali dataset, the recall score of ChatGPT based on \gptfour (0.94) is 32.4\% higher than that of ChatGPT based on \gptthree (0.71). However, the precision score decreases by 52.6\% at the same time. The results indicate that ChatGPT based on \gptfour can report more security-related bug reports with a relatively high false positive rate. 
In practice, this means software maintainers need to spend more time manually checking false positives. With a high recall score and a low precision score, ChatGPT can still benefit the prediction of security bug reports since (1) missing a security-related bug report can lead to severe consequences and (2) it already filters many non-security-related reports. 

\textbf{Implications.} 
When analyzing the results obtained with the \zeroshot prompt, we find that ChatGPT exhibits hallucinations regarding the definition of a security bug report. For instance, ChatGPT incorrectly identifies memory leakage (MML) and null pointer dereference (NPD) as not security-related. 
Providing ChatGPT an MML bug report labeled ``security-related'' in the \oneshot prompt mitigates this issue and improves the the recall score by 1.17 times. However, the precision score decreases due to other hallucinations, i.e., ChatGPT learns some unrelated information from the demonstration example. For instance, when using the \oneshot prompt shown in~\autoref{tab:templateexample2}  (deferred to~\autoref{a:a}), ChatGPT tends to mistakenly mark reports that contain unrelated words of the demonstration example report (e.g., ``tab'' and ``open'') as security-related. 
The results indicate that \textit{when providing demonstration examples, how to make ChatGPT focus on helpful information rather than irrelevant content is an interesting question.} 

Furthermore, in the \manualinfo prompt, we directly tell ChatGPT ``bug reports related to memory leak or null pointer problems should be seen as security bug reports''. Then, for ChatGPT based on \gptthree, this prompt achieves the best performance. The results indicate that the summary of domain knowledge can directly benefit the improvement of ChatGPT's performance in security bug report identification.  

\textbf{Failed Cases Analysis.}
Although the \manualinfo prompt helps mitigate hallucinations, ChatGPT still incorrectly identifies many resource leakage issues as not security-related, causing 17\% of the failed cases. 
This highlights the importance to address this issue in future work.
Additionally, in some failed cases, ChatGPT points out that the description in the bug report is not detailed enough, making it difficult to determine whether it is security-related.
To further understand what kind of information can help ChatGPT with better identification, we ask ChatGPT what additional information it needs. It indicates that \textit{details about the impact and exploitability of bugs are particularly helpful.}

\subsection{Vulnerability Severity Evaluation}
\label{sub-s:severity-result}

In this evaluation, we ask ChatGPT to map a function to the CVSS exploitability metrics~\cite{CVSS} based on its description. 

\textbf{ChatGPT's Performance.} 
As shown in~\autoref{tab:vulnerability-severity}, ChatGPT's performance is slightly inferior to the SOTA approach. 
Specifically, compared to the scores of DiffCVSS, the recall and precision scores of ChatGPT based on \gptfour with the \gptinfo prompt template are 3.5\% and 11.2\%, respectively, lower on average. However, ChatGPT can outperform DiffCVSS on several scores. For instance, when mapping AV:Network, the recall of ChatGPT based on \gptfour with the \gptinfo prompt template (0.9848) is 6.6\% higher than that of DiffCVSS (0.9242). 
Overall, these results show the potential of using ChatGPT in this challenging task. 

\textbf{The Impact of Prompts and Models.} 
Except for the \gptinfo template, 
ChatGPT's performance with other prompt templates is relatively poor in this task. 
We take the results on the PR:High metric as an example to demonstrate this investigation. ChatGPT's recall scores on this metric are 61.0\%, 88.9\%, 72.2\%, 22.0\%, and 44.3\% lower than the SOTA approach (0.9452) when using \zeroshot (0.3684), \oneshot (0.1053), \fewshot (0.2632), \prompteng (0.7368), and \manualinfo templates (0.5263), respectively. However, with the \gptinfo prompt templates, ChatGPT's performance significantly improves. For instance, for \gptthree, compared to the scores with the \manualinfo template, the recall and precision scores with the \gptinfo template improved by 1.60 and 1.73 times on average, respectively.
Besides, the used models also have an essential impact on ChatGPT's performance for this task. With the same \gptinfo prompt, ChatGPT based on \gptfour outperforms the one based on \gptthree on most scores. Specifically, the precision score of the former is 10.5\% higher than the latter on average. 

\begin{table*}
    \small
    \tblcolor
	\centering
	\caption{The evaluation result on vulnerability repair. Gen = Generated. Vld = compilable. Vuln = Vulnerable. Fn = Functional. Safe = Not Vulnerable. Fixed = Fixed Vulnerabilities.  Orig = Using the original code grafting method designed for LLMset~\cite{pearce2021examining}.}
	\label{tab:fixing-code-2}
	\setlength{\tabcolsep}{1.6mm} 
 	{
	\begin{tabular}{cccccccccccccccccccccc}
		\toprule
        \textbf{Approach} & \textbf{Prompt} & \textbf{Dataset} & \textbf{\# Gen} & \textbf{\# Vld} & \textbf{\# Vuln} & \textbf{\# Fn} & \textbf{\makecell[c]{\# Fn \& Vuln}} & \textbf{\makecell[c]{\# Fn \& Safe}} & \textbf{\makecell[c]{\% Vld Repair}} & \textbf{\makecell[c]{\# Fixed}}\\
        \midrule
        \rcolor 
        ExtractFix~\cite{gao2021beyond} & - & \test & - & - & - & - & - & - & - & \textbf{10} \\
        LLMset~\cite{pearce2021examining} & \zeroshot & \test & 3,300 & 674 & 234 & 388 & 252 & 159 & 23.6 & 5 \\
        \rcolor
        LLMset~\cite{pearce2021examining} & \manualinfo & \test & 3,300 & 1254 & 726 & 926 & 705 & 221 & 17.6 & 8 \\
        \midrule
        \gptthree & \zeroshot & \vali & 350 & 329 & 23 & 166 & 5 & 161 & 48.9 & 5 \\
        \rcolor \gptthree & \oneshot &  \vali & 350 & 326 & 8 & 176 & 7 & 169 & 51.8 & 5 \\
        \gptthree & \fewshot & \vali & 350 & 337 & 7 & 145 & 4 & 141 & 41.8 & 6 \\
        \rcolor \gptthree & \prompteng & \vali & 350 & 204 & 4 & 118 & 4 & 114 & 55.9 & 4 \\
        \gptthree (Orig.) & \manualinfo & \vali & 350 & 138 & 40 & 78 & 39 & 39 & 28.3 & 5 \\
        \rcolor \gptthree & \manualinfo & \vali & 350 & 259 & 40 & 227 & 39 & 188 & 72.6 & 7 \\
        \gptthree & \gptinfo & \vali & 350 & 253 & 7 & 153 & 7 & 146 & 57.7 & 6 \\
        \rcolor \gptfour & \manualinfo & \vali & 350 & 292 & 2 & 290 & 2 & 288 & \textbf{98.6} & 7 \\
    \midrule
         \gptfour & \manualinfo & \test & 600 & 377 & 20 & 370 & 20 & 350 & 92.8 & \textbf{10} \\
	\bottomrule
	\end{tabular}}
\end{table*}

\textbf{Implications.} 
In fact, ChatGPT struggles with this challenging task. At the beginning of this evaluation, we find that ChatGPT performs poorly with the \zeroshot, \oneshot, and \fewshot templates. Thus, we attempt to improve its performance in many ways. First, we try to provide more demonstration examples with \fewshot prompts. However, due to the maximum token limitation (8,192 tokens for each question and answer), providing many demonstration examples within one prompt is impractical. 
Inspired by ChatGPT's outstanding NLP capability, we try to compress the demonstration examples with ChatGPT. Consequently, ChatGPT can reduce the length of the demonstration examples by half on average. In this way, we can provide more demonstration examples in one prompt. However, we find that the results remain unsatisfactory. 

Furthermore, we notice that providing the description of CVSS metrics in the \manualinfo template slightly improves the performance. Thus, we try to manually adjust the description in the \manualinfo template. Unfortunately, it is hard to construct perfect expertise content for this task. 
After multiple failures in summarizing the expertise manually, we turn to explore whether ChatGPT can assist us in generating expertise. 
Specifically, we provide the compressed demonstration examples to ChatGPT and ask it to summarize the characteristics of each CVSS metric.~\autoref{lst:GPTknowledge} shows the knowledge summarized by ChatGPT. 
Then, we integrate the summarized knowledge in the \gptinfo prompt. 
As shown in~\autoref{tab:vulnerability-severity}, this \gptinfo prompt improves ChatGPT's performance significantly. 
Hence, we conclude that \textit{extracting expertise by leveraging ChatGPT and incorporating the extracted expertise into prompt is an interesting future research direction, offering the potential to achieve strong performance in challenging vulnerability management tasks.} 

\textbf{Failed Cases Analysis.}
In this task, we identify two types of failed cases. The first type is due to the inherent ambiguity of the CVSS itself. For example, AV:Network and AV:Adjacent are both related to networks, which makes it challenging for ChatGPT to select the appropriate metric. Similarly, in the AC and PR metrics, overlapping characteristics related to check conditions can lead to confusion. The second type of failure arises from our rudimentary implementation of the self-heuristic prompt template. For some metrics, our limited sample diversity hinders ChatGPT from forming a well-rounded summary of expert knowledge. For instance, in the self-heuristic samples for the AC:High metric, we overlook the function about checking flags, which leads ChatGPT to fail in summarizing this characteristic and consequently results in incorrect outcomes. Therefore, \textit{when leveraging a self-heuristic prompt template, it is valuable to explore how to extract a variety of samples and guide ChatGPT to make a more comprehensive summary.}

\begin{figure}[h]
\begin{lstlisting}[caption={},label={}]
Network: Functions that involve network communication, socket handling, or network device management. Examples: sock_register, udp4_hwcsum, ... 
Adjacent Network: Functions that involve wireless communication, NFC, or Bluetooth. Examples: nfc_start_poll, lib80211_wep_encrypt, ... 
Physical: Functions that involve hardware interaction, device management, or USB handling. Examples: usb_release_dev, snd_card_free, ... 
Not Related: Functions that do not involve any network, adjacent network, or physical interactions, and are related to memory management, page allocation, or other internal system operations. Examples: do_set_mempolicy, do_page_mkwrite, ...
\end{lstlisting}\vspace{-8pt}
\caption{An example of the knowledge summarized by ChatGPT.}
\label{lst:GPTknowledge}
\end{figure}

\subsection{Vulnerability Repair}
\label{sub-s:patch-generation-result}
In this evaluation, we ask ChatGPT to fix vulnerabilities in the provided code. The baseline is the SOTA approach ExtactFix~\cite{gao2021beyond} and six LLMs (named LLMset) evaluated by Pearce et al.~\cite{pearce2021examining}, including OpenAI's Codex models~\cite{Codex} and AI21's Jurassic-1 models~\cite{AI21}. We use the dataset provided by Pearce et al.~\cite{pearce2021examining}, which includes 7 hand-crafted vulnerabilities and 12 real-world CVEs. For each vulnerability, the dataset contains (1) the original buggy file and the vulnerable code snippet, (2) the PoC input that triggered the vulnerability, and (3) the project's regression test.  
With this dataset, we provide the vulnerable code snippet to ChatGPT and ask it to fix the vulnerability. After ChatGPT generates a response, we graft it with the original buggy file. Then, we use the PoC to test whether the vulnerability has been fixed and the regression tests to ensure that the fix does not break other functionality.

\begin{table*}
    \small
    \tblcolor
	\centering
    \caption{The evaluation result on vulnerability repair for each CVE. The results are presented as `\# Fn \& Safe'/`\# Vld'. `\cmark' means that the vulnerability was successfully fixed, while `\xmark' is the opposite.}
	\label{tab:fixing-code-1}
	\setlength{\tabcolsep}{0.9mm} 
 	{
	\begin{tabular}{cccccccccccccccccccccc}
		\toprule
        \textbf{Approach} & \textbf{Prompt} & 
        \makecell[c]{EF01\\CVE-\\2016-\\5321} &
        \makecell[c]{EF02_01\\CVE-\\2014-\\8128} &
        \makecell[c]{EF02_02\\CVE-\\2014-\\8128} &
        \makecell[c]{EF07\\CVE-\\2016-\\10094} &
        \makecell[c]{EF08\\CVE-\\2017-\\7601} &
        \makecell[c]{EF09\\CVE-\\2016-\\3623} &
        \makecell[c]{EF10\\CVE-\\2017-\\7595} &
        \makecell[c]{EF15\\CVE-\\2016-\\1838} &
        \makecell[c]{EF17\\CVE-\\2012-\\5134} &
        \makecell[c]{EF18\\CVE-\\2017-\\5969} &
        \makecell[c]{EF20\\CVE-\\2018-\\19664} &
        \makecell[c]{EF22\\CVE-\\2012-\\2806} \\
        \midrule
        \rcolor 
        ExtractFix~\cite{gao2021beyond} & - & \cmark & \cmark & \xmark & \cmark & \cmark & \cmark & \cmark & \cmark & \cmark & \cmark & \xmark & \cmark \\
        LLMset~\cite{pearce2021examining} & \zeroshot & 33/49 & 0/2 & 0/81 & - & 42/135 & 4/4 & 4/65 & - & 53/58 & 0/13 & 0/198 & 0/69 \\
        \rcolor
        LLMset~\cite{pearce2021examining} & \manualinfo & 14/117 & 23/124 & 0/205 & - & 46/78 & 96/190 & 11/37 & 3/98 & 24/33 & 0/120 & 4/171 & 0/81 \\
    \midrule
        \gptfour & \manualinfo & 31/38 & 50/50 & - & 4/6 & 0/5 & 50/50 & 32/34 & 2/4 & 47/50 & 37/43 & 47/47 & 50/50 \\
    \midrule
        \rcolor
        \gptfour/LLMs/EF & - & 
            \cmark/\cmark/\cmark & \cmark/\cmark/\cmark & \xmark/\xmark/\xmark & 
            \cmark/\xmark/\cmark & \xmark/\cmark/\cmark & \cmark/\cmark/\cmark & 
            \cmark/\cmark/\cmark & \cmark/\cmark/\cmark & \cmark/\cmark/\cmark & 
            \cmark/\xmark/\cmark & \cmark/\cmark/\xmark & \cmark/\xmark/\cmark\\
	\bottomrule
	\end{tabular}}
\end{table*}

\textbf{ChatGPT's Performance.}
LLMset generates 50 responses for each CVE with each LLM (except for AI21, which only generates 300 responses in total due to its comparatively lower API usage limit), resulting in 3,300 generated responses. Similarly, we also generate 50 responses for each CVE with ChatGPT, resulting in 600 responses with each prompt template. 
\autoref{tab:fixing-code-2} summarizes the evaluation results of these responses. 
First, ChatGPT based on \gptfour with the \manualinfo prompt can fix 10 / 12 vulnerabilities, which is better than LLMset. 
Second, we observe that compared to LLMset, ChatGPT can achieve a better valid repair rate (i.e., `\# Fn. \& Sate' / `\# Vld').
Specifically, the valid repair rate of ChatGPT based on \gptfour with the \manualinfo prompt is 92.8\%, which is 4.3 times higher than that of LLMset with the same prompt (17.6\%).
Additionally, considering that hallucinations may generate syntactically correct patches but do not actually fix the vulnerability, we manually checked the generated patches to filter such potential false positives, and the results indicate the absence of such occurrences.

\textbf{The Impact of Prompts and Models.}
In this evaluation, we have the following observations about the impact of prompt templates and models. 
First, we observe that the \oneshot prompt has a negative effect. Specifically, for ChatGPT based on \gptthree, although its valid repair rate with the \oneshot prompt (51.8\%) is higher than that with the \zeroshot prompt (48.9\%), less code generated with the \oneshot prompt is compilable.
Besides, the \oneshot prompt does not help ChatGPT fix more vulnerabilities than the \zeroshot prompt (both of them fix 5 vulnerabilities). 
We also tried to provide more bug patches with the \fewshot prompt. Compared to the \oneshot prompt, the \fewshot prompt helps \gptthree fix one more vulnerability. However, the \fewshot prompt lead to a low valid repair rate. The results indicate that providing existing bug patches to ChatGPT cannot significantly help it generate safe and functional code to fix security vulnerabilities. This observation makes sense because it is hard to fix a vulnerability by following how someone fixes other vulnerabilities since each vulnerable program has its own code semantics.

Second, expert knowledge can help ChatGPT in this hard task. Specifically, the \manualinfo prompt of this task, which provides a comment that describes the vulnerability, e.g.,  
``BUG: stack buffer overflow'', can help ChatGPT fix 2 vulnerabilities.  
Third, \gptfour has a clear advantage over \gptthree for this task. Specifically, ChatGPT based on \gptfour generates more compilable code (292) than that based on \gptthree (253). Meanwhile, the valid repair rate of \gptfour (98.6\%) is significantly higher than that of \gptthree (72.6\%). 

\textbf{Implications.} 
\autoref{tab:fixing-code-1} provides more details about the evaluation results for each tested real-world CVE. From~\autoref{tab:fixing-code-1}, we observe several interesting findings. 
First, ChatGPT can fix three CVEs (EF07, EF18 and EF22) which LLMset could not. The fixes of these CVEs are too onerous for LLMset. For instance, EF18's real-world patch is long, removing 10 lines and adding 14;
Besides, ChatGPT can fix one CVE (EF20) which ExtractFix could not. ExtractFix failed in this case since it cannot extract this vulnerability's crash-free constraint (CFC). In particular, extracting the CFC with traditional program analysis is quite challenging, which ExtractFix also failed to do in this case. 
The results indicate that \textit{ChatGPT can be a great choice when traditional program analysis methods fail.} 

Second, at the beginning of this evaluation, we find that only a few programs generated based on ChatGPT's response are compilable. The reason is that ChatGPT tends to repeat several code lines before the bug location of the original buggy file in its responses. Thus, the method that grafts LLMset's responses (start at bug locations~\cite{pearce2021examining}) with the original buggy file cannot properly graft ChatGPT's response, making the generated file not compilable due to syntax or semantic errors. For instance, we provide the results of the original code grafting method for ChatGPT based on \gptthree with the \manualinfo prompt in~\autoref{tab:fixing-code-2} (marked with ``Orig''). The results reveal that this method cannot generate compilable programs for some tested vulnerabilities.
Hence, we developed a new code grafting method that cuts overlap in ChatGPT's response and the original file before the bug location. After that, we obtain the remaining results shown in~\autoref{tab:fixing-code-2}, which indicate that more grafted programs are compilable. Although the problem is not caused by the inherent limitation of ChatGPT, we learn an interesting finding from it, i.e., \textit{when using ChatGPT in complex tasks, directly finishing the task with ChatGPT's response can be impractical. Developing workflows to process and leverage ChatGPT's responses appropriately is an important research direction.}

\begin{table*}
    \small
    \tblcolor
	\centering
	\caption{The evaluation result on patch correctness assessment (compared with Quatrain~\cite{tian2022change}). }
	\label{tab:patch-correctness}
	\setlength{\tabcolsep}{2.9mm} 
 	{
	\begin{tabular}{cccccccccccccccccccccc}
		\toprule
        \textbf{Approach} & \textbf{Prompt} & \textbf{Dataset} & \textbf{Accuracy} & \textbf{+Recall} & \textbf{-Recall} & \textbf{Precision} & \textbf{F1} & \textbf{AUC}\\
        \midrule
        Quatrain~\cite{tian2022change} & - & \test & 0.775 & 0.786 & 0.773 & 0.371 & 0.504 & \textbf{0.858} \\
        \midrule
        \rcolor \gptthree & \zeroshot & \vali & 0.617 & 0.577 & 0.625 & 0.246 & 0.345 & 0.601 \\ 
        \gptthree & \oneshot & \vali & 0.682 & 0.479 & 0.725 & 0.270 & 0.345 & 0.602 \\ 
        \rcolor\gptthree & \fewshot & \vali & 0.720 & 0.493 & 0.768 & 0.311 & 0.381 & 0.631 \\ 
        \gptthree & \prompteng & \vali & 0.797 & 0.359 & 0.889 & 0.408 & 0.382 & 0.624 \\ 
        \rcolor\gptthree & \manualinfo & \vali & 0.761 & 0.479 & 0.821 & 0.362 & 0.412 & 0.650 \\ 
        \gptthree & \gptinfo & \vali & \textbf{0.837} & 0.366 & \textbf{0.937} & \textbf{0.553} & 0.441 & 0.652 \\ 
        \rcolor\gptfour & \gptinfo & \vali & 0.789 & 0.275 & 0.898 & 0.364 & 0.313 & 0.587 \\ 
    \midrule 
        \gptthree & desc-code & \vali & 0.725 & 0.697 & 0.731 & 0.355 & 0.470 & 0.714 \\ 
        \rcolor \gptthree & code-only & \vali & 0.564 & 0.817 & 0.510 & 0.261 & 0.396 & 0.663 \\ 
        \gptfour & desc-code & \vali & 0.700 & \textbf{0.915} & 0.655 & 0.360 & 0.517 & 0.785 \\ 
        \rcolor \gptfour & code-only & \vali & 0.816 & 0.901 & 0.798 & 0.487 & \textbf{0.632} & 0.850 \\ 
    \midrule 
        \gptfour & code-only & \test & 0.819 & 0.868 & 0.811 & 0.439 & 0.583 & 0.840 \\ 
	\bottomrule
	\end{tabular}
 }
\end{table*}

\textbf{Failed Cases Analysis.}
In this task, ChatGPT fails to repair two vulnerabilities. For vulnerability EF02_02, ChatGPT's repair attempt differs from the developer's method, which causes the bug location method based on the developer's repair position~\cite{pearce2021examining} to be ineffective, making all generated files not compilable. For EF08, the failure is due to insufficient vulnerability-related context provided. In particular, EF08 involves a shift range error, but the context does not include information about the shift variable, leading ChatGPT to ``guess'' the variable name and thereby failing to generate the correct repair code. Therefore, to further improve ChatGPT's vulnerability repairing capability in real-world applications, \textit{it could be effective to apply more advanced program slicing methods to provide specific vulnerability-related context.}

\subsection{Patch Correctness Assessment}
\label{sub-s:patch-correctness-result}

In this evaluation, we ask ChatGPT to determine whether a patch correctly fixes a bug. Three SOTA approaches are used as the baselines in this evaluation. We compare ChatGPT with these baselines separately since they use different datasets. Noting that the dataset of Quatrain~\cite{tian2022change} only contains the description of patches while the datasets of Invalidator~\cite{le2023invalidator} and Panther~\cite{tian2023best} only contain the code of patches. 

\textbf{ChatGPT's Performance.}
The comparison results with three SOTA approaches are shown in \autoref{tab:patch-correctness}, \autoref{tab:patch-correctness-2}, and \autoref{tab:patch-correctness-3}. Generally speaking, with advanced prompts and models, ChatGPT performs comparably to the SOTAs. 
Specifically, ChatGPT based on \gptfour outperforms Invalidator on all metric scores. It also outperforms Quatrain and Panther regarding the F1 score. 
For instance, on the test dataset, the F1 scores of ChatGPT based on \gptfour are 15.7\% and 6.7\% higher than Quatrain and Panther, respectively. 
Moreover, considering the -recall metric that reflects the capability of identifying incorrect patches, \gptfour exceeds all SOTA approaches on the corresponding test dataset. This indicates that \gptfour recognizes more incorrect patches, thereby reducing the risk of security instability. All these results highlight the potential of leveraging ChatGPT to assist maintainers in this task.

\begin{table*}
    \small
    \tblcolor
    \centering
	\caption{The evaluation result on patch correctness assessment (compared with Invalidator~\cite{le2023invalidator}). }
	\label{tab:patch-correctness-2}
	\setlength{\tabcolsep}{2.9mm} 
    {
    \begin{tabular}{ccccccccc}
		\toprule
        \textbf{Approach} & \textbf{Prompt} & \textbf{Dataset} & \textbf{Accuracy} & \textbf{+Recall} & \textbf{-Recall} & \textbf{Precision} & \textbf{F1} & \textbf{AUC}\\
        \midrule
        \rcolor Invalidator~\cite{le2023invalidator} & - & \test & 0.813 & 0.900 & 0.789 & 0.540 & 0.675 & 0.844 \\
    \midrule
        \gptthree & \zeroshot & \vali & 0.568 & 0.758 & 0.415 & 0.510 & 0.610 & 0.586 \\
        \rcolor \gptthree & \oneshot & \vali & 0.581 & \textbf{0.970} & 0.268 & 0.516 & 0.674 & 0.619 \\
        \gptthree & \fewshot & \vali & 0.595 & 0.576 & 0.610 & 0.543 & 0.559 & 0.593 \\
        \rcolor \gptthree & \prompteng & \vali & 0.608 & 0.576 & 0.634 & 0.559 & 0.567 & 0.605 \\
        \gptthree & \manualinfo & \vali & 0.621 & 0.545 & 0.683 & 0.581 & 0.563 & 0.614 \\
        \rcolor \gptthree & \gptinfo & \vali & 0.730 & 0.758 & 0.707 & 0.676 & 0.714 & 0.732 \\
        \gptfour & \gptinfo & \vali & 0.757 & 0.667 & \textbf{0.829} & \textbf{0.759} & 0.710 & 0.748 \\
    \midrule
        \rcolor \gptfour & \gptinfo & \test & \textbf{0.849} & 0.933 & 0.826 & 0.596 & \textbf{0.727} & \textbf{0.880} \\
	\bottomrule
	\end{tabular}
}\vspace{-10pt}
\end{table*}

\begin{table*}
    \small
    \tblcolor
    \centering
	\caption{The evaluation result on patch correctness assessment (compared with Panther~\cite{tian2023best}). }
	\label{tab:patch-correctness-3}
	\setlength{\tabcolsep}{2.9mm} 
 	{
    \begin{tabular}{ccccccccc}
		\toprule
        \textbf{Approach} & \textbf{Prompt} & \textbf{Dataset} & \textbf{Accuracy} & \textbf{+Recall} & \textbf{-Recall} & \textbf{Precision} & \textbf{F1} & \textbf{AUC}\\
    \midrule
        \rcolor  Panther~\cite{tian2023best} & - & \test & 0.745 & 0.811 & 0.670 & 0.738 & 0.773 & 0.818 \\
        \midrule
        \gptthree & \zeroshot & \vali & 0.710 & 0.963 & 0.381 & 0.669 & 0.789 & 0.672 \\ 
        \rcolor \gptthree & \oneshot & \vali & 0.642 & 0.972 & 0.214 & 0.616 & 0.754 & 0.593 \\ 
        \gptthree & \fewshot & \vali & 0.653 & \textbf{0.981} & 0.226 & 0.622 & 0.762 & 0.603 \\ 
        \rcolor \gptthree & \prompteng & \vali & 0.720 & 0.844 & 0.560 & 0.713 & 0.773 & 0.702 \\ 
        \gptthree & \manualinfo & \vali & 0.715 & 0.771 & 0.643 & 0.737 & 0.753 & 0.707 \\ 
        \rcolor \gptthree & \gptinfo & \vali & 0.730 & 0.844 & 0.583 & 0.724 & 0.780 & 0.714 \\
        \gptfour & \gptinfo & \vali & \textbf{0.870} & 0.899 & \textbf{0.833} & \textbf{0.875} & \textbf{0.887} & \textbf{0.866} \\ 
    \midrule
        \rcolor \gptfour & \gptinfo & \test & 0.813 & 0.829 & 0.794 & 0.821 & 0.825 & 0.811 \\ 
	\bottomrule
	\end{tabular}
}\vspace{-10pt}
\end{table*}

\textbf{The Impact of Prompts and Models.}
From~\autoref{tab:patch-correctness}, we observe that ChatGPT cannot effectively determine patch correctness with straightforward prompt templates. For instance, on the \vali dataset, the F1 score of ChatGPT based on \gptthree with the \zeroshot prompt is 31.5\% lower than that of Quatrain.
Compared to the \zeroshot template, advanced prompt templates increase several metric scores while decreasing others. However, we notice that all advanced prompt templates can increase the -recall score. Specifically, for \gptthree, compared to the \zeroshot template (0.625), its -recall score increase 16.0\%, 22.9\%, 42.2\%, 31.4\%, and 49.9\% when using \oneshot (0.725), \fewshot (0.768), \prompteng (0.889), \manualinfo (0.821), and \gptinfo templates (0.937), respectively. These results indicate that advanced prompt templates can benefit the identification of incorrect patches. 

\textbf{Implications.} 
At the beginning of this evaluation, we find that on the dataset of Quatrain, 
ChatGPT cannot perform well with all the prompt templates described in~\autoref{tab:templates}. After analyzing ChatGPT's responses manually, we find that ChatGPT complains that it cannot assess the correctness of patches without code. Hence, we manually collect the corresponding code for each patch and develop a new desc-code template that simultaneously provides the code and description.
As shown in~\autoref{tab:patch-correctness}, ChatGPT with the desc-code template still performs somewhat worse than Quatrain.

After manual analysis, we find out an important reason is that \textit{ChatGPT misuses the information in the prompt}. Specifically, when the code and description are provided simultaneously, ChatGPT tends to analyze whether the code changes match the description rather than the correctness of the patch. For example, ChatGPT tends to report that a patch is incorrect when the patch does not modify a function mentioned in the description. 
Thus, we further develop the code-only prompt template that removes descriptions from the desc-code template. 
As shown in~\autoref{tab:patch-correctness}, when using the code-only template, although the performance of ChatGPT based on \gptthree does not increase significantly, ChatGPT based on \gptfour achieves a comparable performance to Quatrain. 
The results indicate that 
\textit{more information is not always better. Guiding ChatGPT to leverage the information in the prompt in a suitable way is an interesting research direction.
}

\textbf{Failed Cases Analysis.}
When examining false positives, we find that by only analyzing the patch code, ChatGPT often misinterprets the logic error that the patch is supposed to fix. For example, ~\cite{commit-math} is supposed to fix a problem of incorrect intersection selection in polyhedron sets. However, without the information about the root cause and correct logic, ChatGPT mistakenly assumes that any seemingly reasonable modification corrects the underlying logic error. 
This issue is also caused by hallucinations, indicating that in the absence of critical information, ChatGPT focuses solely on the surface aspects of the patch and overlooks the essence of the issue it aims to resolve.
On the other side, false negatives are mainly caused by unconventional repair patches generated by automatic repair tools. Specifically, rather than directly removing a problematic control block, some automatically generated patches alter the condition within the control block to ``if (false)''. Therefore, combining the analysis of the implications, we can conclude that more information is not always better, but professional and accurate information is essential. Additionally, while considering removing potential noise in bug reports, it is also necessary to provide more specialized knowledge, such as vulnerability localization and root cause analysis.

\subsection{Stable Patch Classification}
\label{sub-s:fixing-classify-result}

In this evaluation, we ask ChatGPT whether a given patch (including the description and the code snippet) is a stable patch. 

\textbf{ChatGPT's Performance.}
\autoref{tab:bug-fixing} shows the stable patch classification results, which demonstrate that compared to the SOTA approach, ChatGPT performs slightly worse. 
Specifically, on the test dataset, for ChatGPT based on \gptfour with the \manualinfo prompt, its recall score is 4.7\% higher than that of PatchNet. However, the remaining scores are 15.0\% lower on average.
Thus, exploring advanced prompts is still demanded to improve ChatGPT's performance on this task.

\textbf{The Impact of Prompts and Models.}
Although ChatGPT cannot achieve capability on par with the SOTA approach for this task, advanced prompt templates can improve its performance. 
For example, as shown in~\autoref{tab:bug-fixing}, for ChatGPT based on \gptthree tested on the \vali dataset, its accuracy scores with the \manualinfo (0.762) and \gptinfo (0.646) prompts are 35.6\% and 14.1\% higher than that with the \zeroshot prompt (0.566), respectively. 

In this task, \gptfour and \gptthree each have their advantages. 
When using the same \manualinfo prompt on the \vali dataset, ChatGPT based on \gptfour outperforms ChatGPT based on \gptthree regarding the recall and F1 scores while performing worse on other scores. 
Considering the application scenario, we suggest that \gptfour is more suitable for this task for the following reasons. First, recall is more important than precision since identifying as many bug-related patches as possible is critical to ensure the security of the stable version code. Second, \gptfour achieves a better F1 score, a comprehensive embodiment of precision and recall scores. 

\textbf{Implications.} 
When using the \zeroshot, \oneshot, \fewshot, and \prompteng prompts, the precision scores are close to 0.5 while recall scores are close to 1. 
After manual analysis, we conjecture that the reason is ChatGPT's misunderstanding of what constitutes a stable patch, resulting in a hallucination where any seemingly reasonable patch is assumed stable. By using the \manualinfo prompt with a clear
definition of stable patch (``fixing a problem that causes a build error, an oops, a hang, data corruption, a real security issue, or some `oh, that's not good' issue''~\cite{linux}), ChatGPT's performance improves significantly. 
Thus, mitigating ChatGPT's hallucinations is an important research direction to improve its performance on complex tasks. 

\begin{table}
    \small
    \tblcolor
	\centering
	\caption{The evaluation result on stable patch classification. ACC = Accuracy. P = Precision. R = Recall. }
	\label{tab:bug-fixing}
	\setlength{\tabcolsep}{0.5mm}
 	{
	\begin{tabular}{c@{\hspace{-0.3mm}}ccccccccccccccccccc}
		\toprule
        \textbf{Approach} & \textbf{Prompt} & \textbf{Dataset} & \textbf{ACC} & \textbf{P} & \textbf{R} & \textbf{F1} & \textbf{AUC} \\
        \midrule
        \rcolor \multicolumn{2}{l}{PatchNet~\cite{hoang2021patchnet}\hspace{3mm}-} & \test & \textbf{0.862} & \textbf{0.839} & 0.907 & \textbf{0.871} & \textbf{0.860} \\
        \midrule
        \gptthree & \zeroshot & \vali & 0.566 & 0.564 & 0.995 & 0.720 & 0.508 \\ 
        \rcolor \gptthree & \oneshot & \vali & 0.555 & 0.558 & 0.986 & 0.713 & 0.496 \\ 
        \gptthree & \fewshot & \vali & 0.557 & 0.561 & 0.964 & 0.709 & 0.501 \\
        \rcolor \gptthree & \prompteng & \vali & 0.568 & 0.565 & \textbf{0.996} & 0.721 & 0.510 \\ 
        \gptthree & \manualinfo & \vali & 0.762 & 0.761 & 0.837 & 0.798 & 0.752 \\ 
        \rcolor \gptthree & \gptinfo & \vali & 0.646 & 0.631 & 0.884 & 0.737 & 0.614 \\ 
        \gptfour & \manualinfo & \vali & 0.736 & 0.694 & 0.945 & 0.800 & 0.708 \\
    \midrule
        \rcolor \gptfour & \manualinfo & \test & 0.733 & 0.679 & 0.950 & 0.792 & 0.716 \\
	\bottomrule
	\end{tabular}}
\end{table}

\textbf{Failed Cases Analysis.}
In actual stable patch classification scenarios, the definition of "stable" is more complex,  to some extent exceeding the scope defined in official documentation~\cite{linux}. For example, some patches that fix minor issues such as "rename helper function"~\cite{commit-rename} 
are accepted into the stable branch, whereas some more serious issues like 
"fix memory leak"~\cite{commit-ML} 
are not included. This might be because besides the explicit definition, determining whether a patch is stable also involves considering other deeper factors. Models trained on relevant datasets (i.e., PatchNet~\cite{hoang2021patchnet}) may have acquired deeper knowledge, thus performing better in tasks with complex definitions than ChatGPT. Based on the above observation, \textit{developing methods to extract deep features from existing datasets to augment ChatGPT's vulnerability management capabilities is a promising research direction.}

\subsection{User Study}
\label{subsec:userstudy}

In addition to demonstrating the superior performance of ChatGPT on vulnerability management tasks, it is crucial to assess its effectiveness in real-world scenarios for developers.
Accordingly, we have designed a pilot user study that focuses on the bug report summarization process. 
This study aims to explore the perceptions of various developer groups regarding the bug report summarizations generated by ChatGPT with those produced by the SOTA work, iTAPE~\cite{chen2020stay}.
It is important to note that all bug reports used in this study are sourced from open-source bug-tracking systems, ensuring that no confidential information is compromised. 

\textbf{Experiment Settings.}
We recruit 20 participants and divide them into two groups—experts and intermediates—to explore how the practical value of using ChatGPT for vulnerability management varies with developer experience.
The experts group consists of five engineers from the industry, each with more than five years of experience, whereas the intermediates group includes fifteen maintainers from open-source repositories, each with at least three years of development experience.
We provide each participant with 100 randomly selected bug report pairs from our test set, each pair consisting of the original bug report and its summarization generated by iTAPE and ChatGPT. 
Participants are instructed to evaluate the quality of each summary based on three criteria: (1) Correctness - The accuracy with which the summarization captures the essence of the bug report.
(2) Conciseness - The succinctness of the summarization.
(3) Readability - The fluency and clarity of the summarization.

We also provide the instruction ``\textit{Please rate on the scale of 1 to 5 (1 - Poor, 2 - Marginal, 3 - Acceptable, 4 - Good, 5 - Excellent)}'' to guide the scoring of each report.

\begin{figure}[h]
\centering
\includegraphics[width=1\linewidth]{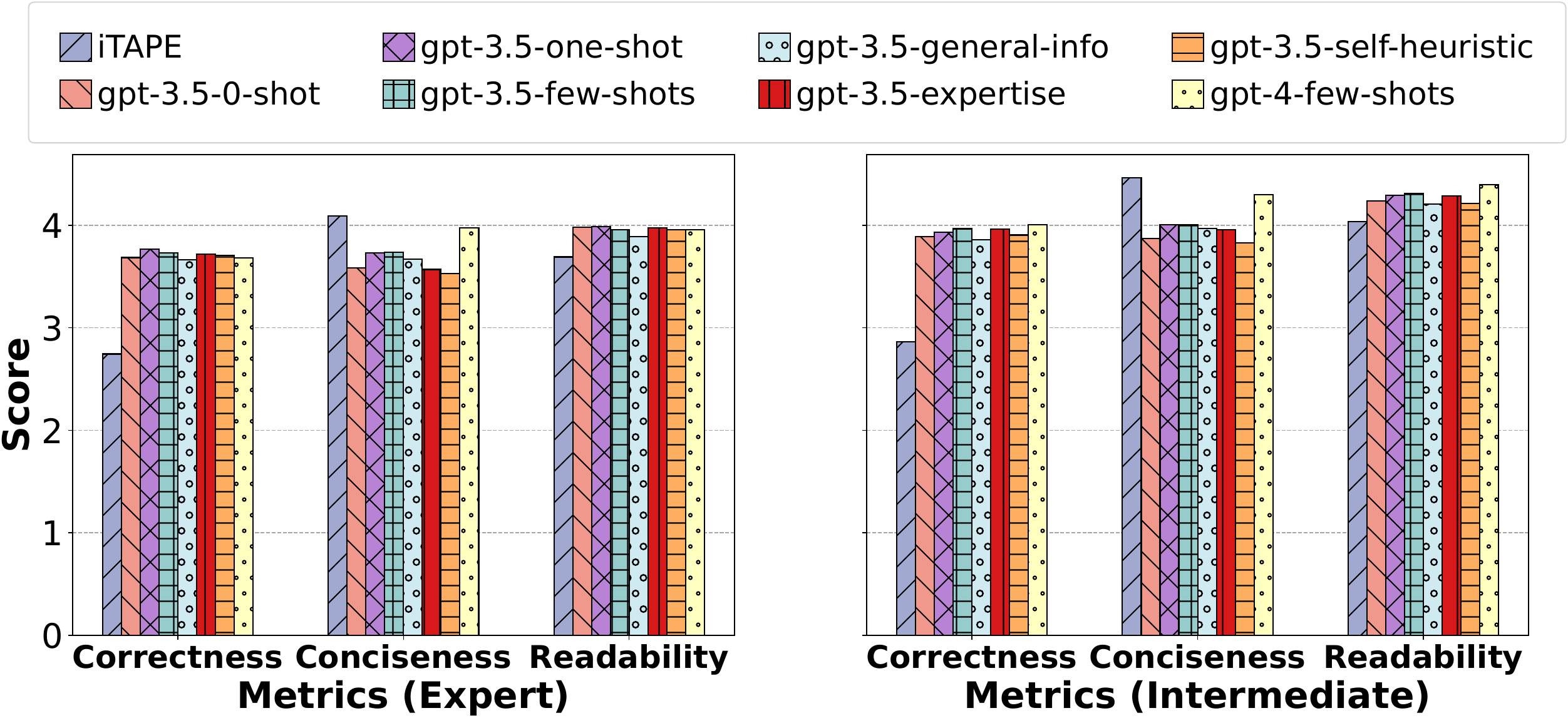} \vspace{-12pt}
\caption{Results of user study (1 - Poor, 2 - Marginal, 3 - Acceptable, 4 - Good, 5 - Excellent).}
\label{fig:user}
\end{figure}

\textbf{Results.}
As depicted in \autoref{fig:user}, the average scores across different dimensions show that ChatGPT generally produces better summarizations in terms of correctness compared to iTAPE, with highest scores of 3.76 (experts) and 4.01 (intermediates), significantly higher than iTAPE's 2.74 (experts) and 2.86 (intermediates).
These results underscore ChatGPT's ability to accurately understand and condense the descriptions of bugs.
However, iTAPE's outputs receive higher conciseness scores than those of ChatGPT.
This outcome is not unexpected, as iTAPE's outputs often omit essential information needed to fully characterize the bug reports.
On average, the bug report summarizations of iTAPE contain 8.5 tokens compared to ChatGPT's 11.3 tokens.
Despite providing more detailed information, ChatGPT's summarizations are rated higher in readability than iTAPE's, suggesting that the additional detail enhances comprehension for developers rather than detracting from it.
Regarding the scores from different participant groups, the professional group generally awards lower scores than the intermediates group, a trend consistent with findings from previous studies \cite{degpt}.
The LLM's training on extensive datasets allows it to perform like a domain expert, which resonates more with intermediate group participants.
However, the professional group's familiarity with the domain means that the incremental information provided by the LLM is less impactful, though they still rate ChatGPT's results well in terms of correctness and readability. Furthermore, no single prompt consistently outperforms others in human evaluations, but \gptfour's outputs tend to be more concise and readable while maintaining high accuracy.
\section{Discussion}
\label{s:discussion}

\textbf{Threats to Validity.}
A potential concern regarding ChatGPT’s performance is test sample leakage during training. Detecting such leakage in LLMs is challenging~\cite{duan2024membership}. Existing membership inference approaches for traditional AI models are difficult to apply to LLMs, requiring impractical conditions such as marked data in the training dataset~\cite{hu2022membership} or surrogate model training~\cite{yang2023gotcha}.  Consequently, existing works infer sample existence by examining match rates between ground truth and LLM responses~\cite{xia2023automated}, which we also adopted. 
Specifically, when summarizing bug reports, if test samples had leaked, ChatGPT would generate numerous summaries similar to the ground truth. However, we observe less than 0.1\% exact matches. To mitigate the potential limitations of exact matching due to ChatGPT's data processing and response randomization, we also investigate fuzzy matches. By leveraging F1 scores under ROUGE-1, we assess the similarity between ground truth and ChatGPT's responses. The results indicate that less than 3.4\% of pairs exhibit a similarity exceeding 70\%. In other tasks, good performance with a straightforward 0-shot prompt would be expected if ChatGPT had seen the test samples, yet the results indicate poor performance. We consider these hypotheses relatively evidential.

Another concern is the potential use of test samples to adjust prompts. To mitigate this, we restricted all prompt template modifications to the training data. Specifically, (1) demonstration examples used in the \oneshot, \fewshot and \gptinfo prompts come from the training dataset, (2) prompt templates are improved based on mistakes observed in training dataset, and (3) the probe-test dataset is separated from the training dataset. Overall, the test dataset remains untouched during prompt adjustment and probe-testing, ensuring a fair evaluation of ChatGPT's capabilities.

\textbf{Limitations and Future Works.} 
\textbf{}(1) \textit{Task complexity}. We do not investigate how ChatGPT's performance scales with task complexity or size, which is regarded as future work.

(2) \textit{Prompting techniques}. We manually constructed prompt templates based on prior works in LLM evaluation~\cite{ma2023scope} 
and our empirical analysis. This manual approach is adopted due to the inherent challenges of automatic prompt engineering (APE), a complex area that holds potential for stimulating research~\cite{shin2020autoprompt}. Moreover, our prompt templates and evaluation pipeline are generally applicable to other LLMs. Addressing adaptability issues is an interesting future research direction.

(3) \textit{Real world interaction scenarios}. Our evaluations are conducted in controlled settings using pre-defined datasets. Exploring scenarios where ChatGPT interacts with actual software development and vulnerability management environments would be insightful.
Additionally, integrating ChatGPT across multiple tasks to reflect the interconnected nature of vulnerability management processes would provide a more comprehensive view of its performance.

(4) \textit{Alternative AI approaches}. We focus on evaluating ChatGPT's performance since it is the most popular AI product nowadays. Conducting further evaluations with other LLMs to investigate and compare their performance is considered as future work. Additionally, exploring alternative AI approaches, such as fine-tuning open-sourced models, may address some identified shortcomings, representing another interesting future research direction.

(5) \textit{Hallucination issues}. We identify several hallucinations and conduct measures such as expertise prompts and manual checks to mitigate this problem. However, addressing hallucinations in LLMs remains an open issue for future research. 

\section{Related Work}
\label{s:relwk}

\textbf{AI for Vulnerability Management.} 
With the rapid development of AI, machine learning models have been enthusiastically promoted as tools for various vulnerability management tasks~\cite{zhou2021spi,li2022dear,lam2015combining,li2019improving}. Some prior works leverage AI methods for automated software artifact generation~\cite{zhang2022itiger}, vulnerability severity examination~\cite{wu2022aware}, and bug fixing~\cite{prenner2022can},  
demonstrating impressive capabilities in these areas. Inspired by this, this paper explores the potential of using ChatGPT, the most emerging AI product currently, for vulnerability management. 

\textbf{The Applications of ChatGPT.} 
ChatGPT has received increasing attention and reputation from various fields. 
In the beginning, ChatGPT is applied in NLP tasks~\cite{deng2022benefits}
and has shown impressive capabilities. 
Considering natural language shares similar aspects as program language, some prior works also explore ChatGPT's capabilities on code analysis~\cite{ma2023scope} and bug fixing~\cite{sobania2023analysis}, indicating the potential for using ChatGPT in software engineering. 
However, can ChatGPT complete vulnerability management tasks that require a systematic understanding of code syntax, program semantics, and related comments? To the best of our knowledge, it is still an unexplored problem. This paper fills this gap by conducting the first large-scale evaluation of ChatGPT's performance for vulnerability management.

\section{Conclusion}
\label{s:conclusion}

This paper conducts the first large-scale evaluation to explore the capabilities of ChatGPT on vulnerability management. Specifically, we compare ChatGPT with 11 SOTA approaches on 6 vulnerability management tasks by using a large-scale dataset containing \numtoken tokens. This systematical investigation allows us to understand the capabilities and limitations of ChatGPT on each task. Our findings demonstrate the desirable prospects of leveraging ChatGPT to assist vulnerability management. Meanwhile, we also reveal the difficulties ChatGPT encountered and shed light on future research to explore better ways to leverage ChatGPT in vulnerability management tasks. 
\section*{Acknowledgments}


This work was partly supported by NSFC under Grant No. 62302443, the Fellowship of China National Postdoctoral Program for Innovative Talents (BX20230307), the Fundamental Research Funds for the Central Universities (Zhejiang University NGICS Platform+226-2024-00048), Jianghuai Advance Technology Center under No. 00QK0021, and Ant Group. Kangjie Lu was supported in part by the NSF awards CNS2045478, CNS-2106771, CNS-2154989, and CNS-2247434. Any opinions, findings, conclusions or recommendations expressed in this material are those of the author and do not necessarily reflect the views of NSF. 



\bibliographystyle{plain}
\bibliography{p}

\appendix
\section{Appendix}
\label{a:a}

\vspace{-6pt}
\begin{figure}[H]
    \centering
    \includegraphics[width=0.475\textwidth]{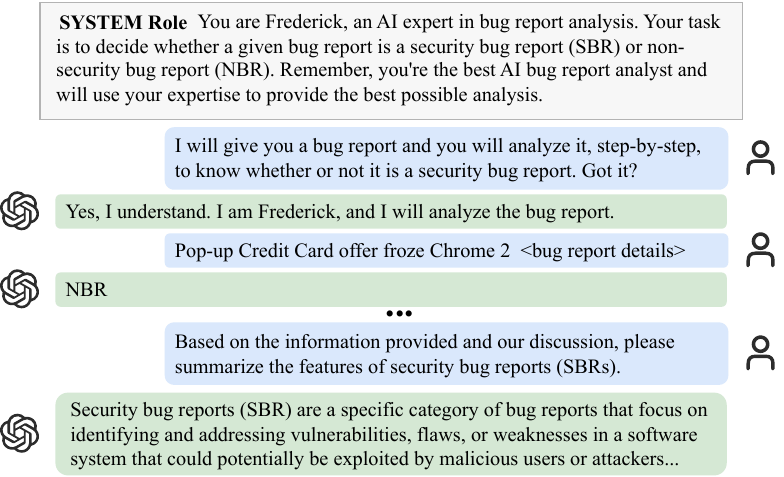} \vspace{-6pt}
    \caption{Extracting expertise from demonstration examples.}
    \label{fig:selfprompt}
\end{figure}

\textbf{Self-heuristic Prompt Generation}. To improve ChatGPT's performance in intricate tasks demanding domain knowledge, such as vulnerability severity evaluation, we introduce the self-heuristic prompt. First, we employ ChatGPT to extract expertise from demonstration examples, as shown in \autoref{fig:selfprompt}. Then, we incorporate this expertise into self-heuristic prompts. 

\vspace{-6pt}
\begin{table}[H]
    \small
    \centering
    \caption{Prompt skills for the \prompteng template.} 
    \label{tab:skill}
    \setlength{\tabcolsep}{3pt}
    \begin{tabularx}{\columnwidth}{>{\centering\arraybackslash}m{2.2cm} >{\raggedright\arraybackslash}X}
        \toprule
        \textbf{Skill} & \textbf{Description} \tabularnewline
        \midrule
        \role & Ask the model to adopt the task-related role. \tabularnewline
        \reinforce & Repeat essential elements of the instructions. \tabularnewline
        \confirmation & Simulate a task confirmation dialogue. \tabularnewline
        \pos & Provide positive feedback prior to the query. \tabularnewline
        \zerocot & Ask the model to think step-by-step. \tabularnewline
        \conclusion & Ask the model to reach the right conclusion. \tabularnewline
        \bottomrule
    \end{tabularx}
\end{table}

\vspace{-4pt}
\textbf{Skills for the General-info Prompt}.~\autoref{tab:skill} provides all prompt skills along with their descriptions used for the \prompteng template. These skills have shown superiority in traditional NLP tasks~\cite{Prompt}.

\vspace{19pt}

\begin{table*}[!b]
    \small 
    \centering    
    \caption{Templates for task prompt generation.} 
    \label{tab:templateexample2}
    \begin{tabularx}{\textwidth}{>{\centering\arraybackslash}m{1.1cm}>{\raggedright\arraybackslash}X}
        \toprule
        \textbf{Name} & \textbf{Template \& Example} \\
        \midrule
            & \rolestyle{USER} <task description> <input> \\
            \cline{2-2}\vspace{4pt}
            \multirow{-1.5}{*}{\zeroshot} 
            & \rolestyle{USER} Decide whether a bug report is a security bug report (SBR) or non-security bug report (NBR). Bug report: <bug report> Category: \\
        \midrule
            & \rolestyle{USER} <task description> \textcolor{oneshotcolor}{<demonstration>} <input> \\
            \cline{2-2}\vspace{4pt}
            \multirow{-1}{*}{\oneshot} 
            & \rolestyle{USER} Decide whether a bug report is a security bug report (SBR) or non-security bug report (NBR). \textcolor{oneshotcolor}{Bug report: Memory Leak in about: memory 1. Open a new tab and enter about:memory... Category: SBR} \#\#\# Bug report: <bug report> Category: \\
        \midrule
            & \rolestyle{USER} <task description> \textcolor{oneshotcolor}{<demonstration 1> <demonstration 2> <demonstration 3> <demonstration 4>} <input> \\
            \cline{2-2}\vspace{4pt}
            \multirow{-1}{*}{few-shot} 
            & \rolestyle{USER} Decide whether a bug report is a security bug report (SBR) or non-security bug report (NBR). \newline \textcolor{oneshotcolor}{Bug report: Mem leak with IPC::Channel::Channel() in unit_tests UtilityProcessHostTest... Category: NBR \#\#\# Bug report: ... Category: NBR \#\#\# Bug report: ... Category: NBR \#\#\# ... } \#\#\# Bug report: <bug report> Category: \\
        \midrule
            & \rolestyle{SYSTEM} \textcolor{role}{<\role>} <task description> \textcolor{encourage}{<\reinforce>} \quad \rolestyle{USER} <task description> \textcolor{mock}{<\confirmation>} \newline
            \rolestyle{ASSISTANT} \textcolor{mock}{<\confirmation>} \qquad \rolestyle{USER} \textcolor{encourage}{<positive feedback>} <input> \textcolor{CoT}{<\zerocot>} \textcolor{encourage}{<\conclusion>} \\
            \cline{2-2}\vspace{4pt}
            \multirow{-1}{*}{\shortstack{general\\-info}} 
            & \rolestyle{SYSTEM} \textcolor{role}{You are Frederick, an AI expert in bug report analysis.} Your task is to decide... \textcolor{encourage}{Remember...} \quad \newline
            \rolestyle{USER} I will give you a bug report and you will... \textcolor{mock}{Got it?} \quad 
            \rolestyle{ASSISTANT} \textcolor{mock}{Yes, I understand. I am Frederick, and I will analyze the bug report.} \quad 
            \rolestyle{USER} \textcolor{encourage}{Great! Let's begin then :)} For the bug report: <bug report> Is this bug report (A) a security bug report (SBR), or (B) a non-security bug report (NBR). Answer: Let's think \textcolor{CoT}{step-by-step} to \textcolor{encourage}{reach the right conclusion}, \\
        \midrule
            & \rolestyle{SYSTEM} \textcolor{role}{<\role>} <task description> \textcolor{encourage}{<\reinforce>} \quad \rolestyle{USER} \textcolor{infocolor}{<expertise>} <task description> \textcolor{mock}{<\confirmation>} \newline
            \rolestyle{ASSISTANT} \textcolor{mock}{<\confirmation>} \qquad \rolestyle{USER} \textcolor{encourage}{<positive feedback>} <input> \textcolor{CoT}{<\zerocot>} \textcolor{encourage}{<\conclusion>} \\
            \cline{2-2}\vspace{4pt}
            \multirow{-1}{*}{\manualinfo} 
            & \rolestyle{SYSTEM} \textcolor{role}{You are Frederick...} Your task is... \textcolor{infocolor}{memory leak or null pointer...} \textcolor{encourage}{Remember...} \quad 
            \rolestyle{USER} \textcolor{infocolor}{A security bug report is...} I will give you a bug report... \textcolor{mock}{Got it?} \quad 
            \rolestyle{ASSISTANT} \textcolor{mock}{Yes, I understand. I am Frederick, and I will analyze the bug report.} \quad 
            \rolestyle{USER} \textcolor{encourage}{Great! Let's begin then :)} For the bug report: <bug report> ... Let's think \textcolor{CoT}{step-by-step} to \textcolor{encourage}{reach the right conclusion}, \\
        \midrule
            & \rolestyle{SYSTEM} \textcolor{role}{<\role>} <task description> \textcolor{encourage}{<\reinforce>} \quad \rolestyle{USER} \textcolor{infocolor}{<knwoledge>} <task description> \textcolor{mock}{<\confirmation>} \newline 
            \rolestyle{ASSISTANT} \textcolor{mock}{<\confirmation>} \qquad \rolestyle{USER} \textcolor{encourage}{<positive feedback>} <input> \textcolor{CoT}{<\zerocot>} \textcolor{encourage}{<\conclusion>} \\
            \cline{2-2}\vspace{4pt}
            \multirow{-2}{*}{\shortstack{self\\-heuristic}} 
            & \rolestyle{SYSTEM} \textcolor{role}{You are Frederick...} Your task is... \textcolor{encourage}{Remember...} \quad 
            \rolestyle{USER} \textcolor{infocolor}{Security bug reports (SBR) are...} I will ... \textcolor{mock}{Got it?} \quad           
            \rolestyle{ASSISTANT} \textcolor{mock}{Yes, I understand...} \quad \rolestyle{USER} \textcolor{encourage}{Great...} For the bug report:... Let's think \textcolor{CoT}{step-by-step} to \textcolor{encourage}{reach the right conclusion}, \tabularnewline
        \bottomrule
    \end{tabularx}
\end{table*}

\newpage

\begin{table}[H]
    \small
    \centering
    \vspace{-8pt}
    \caption{Task-specific expertise and the referred resources.} 
    \label{tab:expertise}    
    \setlength{\tabcolsep}{0.6pt}
    \begin{tabularx}{\columnwidth}{>{\centering\arraybackslash}m{1.7cm} >{\raggedright\arraybackslash}m{\columnwidth-1.7cm-2.4pt}}
        \toprule
        \textbf{Task} & \textbf{Expertise} \\
        \midrule
        Bug report summarization & "1. The titles should be within the range of 5 to 15 words, and not contain URLs. 2. At least 30\% words of the titles should come from the bug report."~\cite{chen2020stay} \\
        \midrule
        Security bug report identification & "When analyzing the bug report, take into account that bug reports related to memory leak or null pointer problems should be seen as security bug report."~\cite{wu2021data} \\
        \midrule
        Vulnerability severity evaluation & "CVSS AV metric: 1. Network: The vulnerable component is bound to the network stack and... 2. Adjacent:...attack must be launched from the same shared physical (e.g., Bluetooth)... 3.Physical:..."~\cite{CVSS} \\
        \midrule
        Vulnerability repair & "[ERROR_MESSAGE], [VULNERABLE_CODE]." \textit{Note: ERROR_MESSAGE and VULNERABLE_CODE differ by vulnerability.} \cite{pearce2021examining} \\
        \midrule
        Patch correctness assessment & "A correct patch implements changes that “answer” to a problem posed by bug report..."~\cite{tian2022change} \\
        \midrule
        Stable patch classification & "The patch accepted in Linux-stable release must fix a real and critical bug that causes a build error, an oops, a hang, ..."~\cite{linux}
        \tabularnewline
        \bottomrule
    \end{tabularx}
\end{table}

\end{document}